\documentclass[fleqn,aps,12pt,preprint,groupedaddress,showpacs,reprintnumbers,amsmath,amssymb,floatfix,nofootinbib]{revtex4-1}

\usepackage{graphicx}
\usepackage{epsfig}
\usepackage{graphics}
\usepackage{latexsym}
\usepackage{amsmath}
\usepackage{amssymb}
\usepackage{rotating}
\usepackage{bm}
\usepackage{color}

\begin{document}

\title[Parameter Optimisation for the Latest QMC Energy Density Functional]{Parameter Optimisation for the Latest Quark-Meson Coupling Energy Density Functional}

\author{K. L. Martinez}
\author{A. W. Thomas}
\address{CSSM and CoEPP, Department of Physics, University of Adelaide, SA 5005 Australia}

\author{J. R. Stone}
\address{Department of Physics (Astro), University of Oxford, OX1 3RH United Kingdom}
\address{Department of Physics and Astronomy, University of Tennessee, TN 37996 USA}

\author{P. A. M. Guichon}
\address{IRFU-CEA, Universit\'e Paris-Saclay, F91191 Gif sur Yvette, France}

\date{\today}

\begin{abstract}
The Quark--Meson--Coupling (QMC) model self-consistently relates the dynamics of the internal quark structure of a hadron to the relativistic mean fields arising in nuclear matter. It offers a natural explanation to some open questions in nuclear theory, including the origin of many-body nuclear forces and their saturation, the spin-orbit interaction and properties of hadronic matter at a wide range of densities. 
 The QMC energy density functionals QMC-I and QMC$\pi$-I  have been successfully applied to calculate ground state observables of finite nuclei in the Hartree-Fock + BCS approximation, as well as to predict properties of dense nuclear matter and cold non-rotating neutron stars. Here we report the latest development of the model, QMC$\pi$-II, extended to include higher order self-interaction of the $\sigma$ meson. A derivative-free optimization algorithm has been employed to determine a new set of the model parameters and their statistics, including errors and correlations. QMC$\pi$-II predictions for a wide range of properties of even-even nuclei across the nuclear chart, with fewer adjustable parameters, are comparable with other models. Predictions of ground state binding energies of even-even isotopes of superheavy elements with Z$>$96 are particularly encouraging.

\end{abstract}


\maketitle

\section{Introduction}
The QMC model is based on the self-consistent adjustment of the internal structure of hadrons immersed in a nuclear medium in which there are strong Lorentz scalar and vector  mean fields \cite{Guichon1988,Guichon1995}. These changes lead naturally to the appearance of many-body nuclear forces through higher order terms in density in the QMC energy density functional (EDF)~\cite{Guichon:2004xg,Guichon:2006er}. The model has been applied to a wide range of problems of experimental interest~\cite{Saito:2005rv}, including the possible existence of meson-nucleus bound states~\cite{Tsushima:1998qw,Bass:2005hn} and the structure of hypernuclei~\cite{Guichon:2008zz}.  

The first systematic application of the QMC model in the Hartree-Fock (HF)+BCS framework to a wide range of even-even nuclei (QMC-I) ~\cite{Stone2016} produced promising results with fewer and well constrained free parameters as compared to the traditional and frequently used EDF of the Skyrme type. The accuracy with which the ground state binding energies of superheavy nuclei were reproduced, although they were not included the fitting procedure, was particularly encouraging. This feature was explored further in \cite{Stone2017}, using the next development of the model by including one-pion-exchange (QMC$\pi$-I).

In this work, we report the latest version of the QMC model (QMC$\pi$-II), further extended by higher order self-interaction of the $\sigma$ meson ~\cite{Guichon2018}. Modern search procedures were used to optimise the model parameters to give the best fit to known ground state properties of a large number of magic and semi-magic nuclei, while retaining consistency with empirical nuclear matter properties at saturation. 

This paper is organized as follows: Sec.~\ref{theory} gives a short outline of the main features of the QMC EDF; the method of obtaining the QMC$\pi$-II parameter set as well the statistics necessary to validate the results against experiment are presented in Sec.~\ref{method}; Sec.~\ref{results} contains assessment of the quality of the fit and summarizes the main results together with their analysis and discussion, followed by Sec.~\ref{conclude} with main conclusions and outlook for future study.

\section{Theoretical framework}
\label{theory}
\subsection{The QMC$\pi$-II EDF}\label{theory:QMC}

The full derivation of the EDF can be found in the recent review \cite{Guichon2018}. Here we outline only the main features of the model. 

The basic idea of the QMC model is to apply self-consistently a scalar mean field $\sigma$ to a bound nucleon which has internal structure, responsive to the effect of the external field. While the structure of the intermediate range attraction between two nucleons is undoubtedly complex, involving various types of two-pion exchange as well as the exchange of the observed $\sigma$ meson, relativistic models have enjoyed considerable success replacing all of this by the exchange of effective mesons ($\sigma, \omega, \rho$). This approach is adopted in the QMC model.

Assuming that the nucleon is modeled as the MIT bag, the equation of motion of the bag in the external fields can be derived.  Solution of the equation of motion of a bag in a constant field yields an effective mass which can be approximated as
\begin{equation}
M^{\text{*}}_B = M_B-g_\sigma\sigma + \frac{d}{2}( g_\sigma\sigma)^2,
\end{equation}
where $g_\sigma\sigma$ is the strength of the scalar field, $g_\sigma$ is the coupling of the scalar meson to the free nucleon and is related to the quark-meson coupling, and $d$ is the scalar polarizability. It quantifies the effects of the scalar field on the nucleon structure and is determined within the model to a good approximation as $d \approx 0.18 R_B$ with $R_B$ being the bag radius. The coupling of the nucleon with the vector meson fields, $g_\omega\omega$ and $g_\rho\rho$, do not affect the internal structure of the bag but contribute a constant shift to its energy.

When applied to finite nuclei, we consider a bag in meson fields which vary slowly as a function of position and assume that in the nucleus the meson fields essentially follow the nuclear density.  This assumption should secure the instantaneous adjustment of the motion of the quarks, which are relativistic, to the actual value of the field.

Taking the nuclear system as a collection of non-overlapping bags, the \textit{classical} total energy can be written as a sum of contributions from the bag motion and the meson fields \cite{Guichon2018}
\begin{equation}
E_{QMC}=\sum_{i=1,...}\sqrt{P_{i}^{2}+M_{i}^{2}(\sigma(\vec{R}_{i})}+g_{\omega}^{i}\omega(\vec{R}_{i})+g_{\rho}\vec{I}_{i}.\vec{B}(\vec{R}_{i})+E_{\sigma}+E_{\omega,\rho},\label{eq:E_QMC}
\end{equation}
where $\vec{R}_{i}$ and  $\vec{P}_{i}$ are the position and momentum of a baryon $i$ and $\vec{I}$ is the isospin matrix. Following the notation of Ref.~\cite{Guichon2018}, $\vec{B}$ stands here for the isovector $\rho$ field to avoid a confusion with the baryon number density $\rho$ used in Sec.~\ref{nmp}.

The energy of the $\sigma$ field is
\begin{equation}
\label{eq:energy}
E_{\sigma}=\int d\vec{r}\left[\frac{1}{2}\left(\vec{\nabla}\sigma\right)^{2}+V(\sigma)\right]
\end{equation}
and the expressions for $E_{\omega,\rho}$ are analogous. 
The potential energy in (\ref{eq:energy}), $V(\sigma)=m_{\sigma}^{2}\sigma^{2}/2+\cdots$, 
is generally limited to the quadratic term. This was the case in the QMC-I and QMC$\pi$-I models.
Here we take a more general form \cite{Guichon2018}
\begin{equation}
\label{eq:Vsigma}
V(\sigma) = \frac{m_{\sigma}^{2}\sigma^{2}}{2}+\frac{\lambda_{3}}{3!}\left(g_{\sigma}\sigma\right)^{3}+\frac{\lambda_{4}}{4!}\left(g_{\sigma}\sigma\right)^{4}.
\end{equation}
This extension involves an additional parameter $\lambda_3$ which must be obtained from a fit to experimental data. The quartic term is added to guarantee the existence of a ground state. The constant $\lambda_{4}$ may be arbitrarily small but must be positive. It has been set to zero in the present work because we are not concerned with the limit of large $g_\sigma \sigma$. The generalization allows a contribution of the $\sigma$-exchange in the t- channel to the polarizability, that cannot arise from the response of the bag to the $\sigma$ field. This extension leads to a significant improvement of the QMC$\pi$-II predictions of saturation properties of symmetric nuclear matter.

After quantization, the Hamiltonian $H_{QMC}$, corresponding to the classical energy (\ref{eq:E_QMC}), still depends on the meson fields. They are eliminated through the equations of motion:
\begin{equation}
\frac{\delta H_{QMC}}{\delta\sigma(\vec{r})}=\frac{\delta H_{QMC}}{\delta\omega(\vec{r})}=\frac{\delta H_{QMC}}{\delta B(\vec{r})}=0 \, ,
\end{equation}
where the $\rho$ field is denoted $B$ to avoid confusion with the density.
In practice, we write the meson field operator $\sigma$=$\bar{\sigma}$ + $\delta\sigma$ (and similarly for the other mesons) where $\bar{\sigma}$ is the expectation value $\langle\sigma\rangle$ determined by the mean field equation
\begin{equation}
\frac{\langle\delta H_{QMC}\rangle}{\delta\bar\sigma} = 0
\end{equation}
and the $\delta\sigma$ is a fluctuation determined as a perturbation around the mean field. In our HF approximation, the fluctuations generate the Fock term. The full QMC Hamiltonian reads
\begin{equation}
H_{QMC}=H_\sigma + H_\omega + H_\rho + H_{so}+H_\pi \, .
\label{eq:Hqmc}
\end{equation}
The first three terms are spin independent. The spin-orbit term, $H_{so}$, arises naturally within the model from the first order correction associated  with the variation of the external field over the volume of the nucleon (see for details section 2.2.2 of Ref.~\cite{Guichon2018}). It is fully expressed in terms of the existing QMC parameters. The pion exchange, because of its long range character is calculated as a perturbation in a local density approximation \cite{Guichon2018} and does not introduce additional free parameters.

The Hamiltonian (\ref{eq:Hqmc})  is used to develop the QMC EDF in HF calculation of finite nuclei 
\begin{equation}
\mathcal{E}_{QMC}= \langle\Phi|H_{QMC}|\Phi\rangle.
\label{eq:EDF}
\end{equation}
The expectation value of $\langle H_{QMC}\rangle$ in a Slater determinant $\Phi$ for Z protons and N neutrons, obtained by filling the single-particle states $\{\phi^{i}(\vec{r},\sigma,m)\}$ up to a Fermi level $F_{m}$ with  $m=\pm1/2$ being the isospin projection such that $p\leftrightarrow1/2,n\leftrightarrow-1/2$, is calculated as a function of density $\rho$, kinetic energy density $\tau$ and spin-orbit density $\vec{J}$
\begin{eqnarray}
\rho_{m}(\vec{r}) & = & \sum_{i\in F_{m}}\sum_{\sigma}\left|\phi^{i}(\vec{r},\sigma,m)\right|^{2},\,\,\,\rho=\rho_{p}+\rho_{n},\\
\tau_{m}(\vec{r}) & = & \sum_{i\in F_{m}}\sum_{\sigma}\left|\vec{\nabla}\phi^{i*}(\vec{r},\sigma,m)\right|^{2},\,\,\,\tau=\tau_{p}+\tau_{n},\\
\vec{J}_{m} & = & i\,\sum_{i\in F_{m}}\sum_{\sigma\sigma'}\vec{\sigma}_{\sigma'\sigma}\times\left[\vec{\nabla}\phi^{i}(\vec{r},\sigma,m)\right]\phi^{i*}(\vec{r},\sigma',m),\,\,\,\vec{J}=\vec{J}_{p}+\vec{J}_{n} \, 
\end{eqnarray}

\subsection{Pairing and Coulomb terms}

Pairing and Coulomb terms are not included in the QMC model. For modeling of finite nuclei, QMC$\pi$-II EDF is augmented by $\mathcal{E}_{\rm pair}$, based on the BCS model with $\delta$-function pairing interaction acting through the whole nuclear volume \cite{Stone2007}
\begin{equation}
\mathcal{E}_{\rm pair} = \frac{1}{4}\sum\limits_{q\in(p,n)}{V^{\rm pair}_q \int{d^3r\chi^2_q}}, \,\,\,\,\,\, \chi_{\rm q}(\vec{r})=
\sum\limits_{\alpha\in{q}} {u{_\alpha}v{_\alpha}|\phi_{\alpha}(\vec{r})|^2},
\label{eq:pair}
\end{equation}
where $q\in(p,n)$, $v_\alpha$, $u_\alpha=\sqrt{1-v_\alpha^2}$ are the occupation amplitudes and $\alpha$ stands for quantum numbers of a single-particle state. The two pairing strengths $V_p^{pair}$ and $V_n^{pair}$ for proton and neutron are two additional parameters to be fitted to experimental data. 

The Coulomb term is taken in its standard form \cite{Klup2009}
\begin{equation}
\mathcal{E}_{\rm Coulomb}=e^2\frac{1}{2}\int{d^3r d^3r^\prime\frac{\rho_p(\vec{r}) \rho_p(\vec{r} ^\prime)}{|\vec{r}-\vec{r}^\prime|}} 
-\frac{3}{4}e^2\Big{(}\frac{3}{\pi}\Big{)}^{\frac{1}{3}}\int{d^3r[\rho_p]^{4/3}},
\end{equation} 
including the exchange term in the Slater approximation. $\rho_p$ stands for density distribution of point-like protons.

\section{Method}
\label{method}
The HF+ BCS calculation was performed using the computer code code SKYAX, allowing for axially symmetric and reflection-asymmetric shapes, adapted by P.-G. Reinhard \cite{skyax,Stone2016} for use with QMC-type EDF. The minimization process was performed in two ways, either without any additional constraint of the path to the final minimum or applying a constraint (CHF) requiring a fixed value of quadrupole moment $< Q_2 >=\frac{3}{4\pi}AR_0^2\beta_2$ with A being a mass number and $R_0$ = 1.2 fm. The latter procedure, particularly useful for calculation of ground state shapes of axially deformed nuclei in terms of the quadrupole deformation parameter $\beta_2$, involves determination of the equilibrium wavefunctions and single-particle energies at each chosen value of $\beta_2$ used to calculate the quadrupole moment. Changing the deformation parameter by a fixed amount through an expected range of deformations yields the lowest energy of the system and its equilibrium shape.

The QMC$\pi$-II EDF depends on three effective coupling constants $G_{\sigma}$, $G_{\omega}$, and $G_{\rho}$
\begin{equation}
G_{\sigma}=\frac{g_{\sigma}^{2}}{m_{\sigma}^{2}},\,\,\, G_{\omega}=\frac{g_{\omega}^{2}}{m_{\omega}^{2}},\,\,\, G_{\rho}=\frac{g_{\rho}^{2}}{m_{\rho}^{2}},
\end{equation}
the $\sigma$ meson mass $m_\sigma$, and the $\sigma$ self-interaction parameter $\lambda_3$. With the two pairing strength  $V_p^{pair}$ and $V_n^{pair}$, there are seven free parameters that need to be fitted to experimental data. The remaining parameters of the model, the $\omega$ and $\rho$ meson masses, and the isoscalar and isovector magnetic moments, which appear in the spin-orbit interaction \cite{Guichon2018}, were taken at their physical values. The MIT bag radius $R_B$ was set to 1 fm.

The fit has been performed first to properties of infinite nuclear matter and further narrowed down using extensive data on ground state observables of even-even finite nuclei. 

\subsection{Nuclear matter properties (NMP)}
\label{nmp}
The EDF (\ref{eq:EDF}) significantly simplifies in infinite nuclear matter, a medium with uniform density $\rho$ without surface and spin-orbit effects. All gradient terms vanish and $\langle H_{QMC} \rangle$ reduces to $\langle H_{NM} \rangle$. The binding energy per particle of cold matter containing protons and neutrons is expressed as a function of density and the proton-neutron ratio
\begin{equation}
\frac{E}{A}(\rho,I)=\frac{\langle H_{NM} \rangle}{\rho}(\rho,I),
\label{eos}
 \end{equation}
where $\rho=\rho_p+\rho_n$ is the total density and $\rho_{p,n}$ are proton and neutron number densities. The neutron excess $I$ is defined as the ratio of the difference between the number of neutrons $N$ and protons $Z$ to the mass number $A$ of the nucleus, $I=(N-Z)/A$.

Symmetric nuclear matter (SNM), with $N$ equal to $Z$ and thus $I=0$, is bound at the saturation point $\rho_0$ $\sim$ 0.16 fm$^{-3}$ fm with energy $E_0\sim$--16 MeV. It is customary to use properties of the SNM at saturation, derived from E/A at $\rho_0$ to constrain parameters of nuclear structure models. In this work we employ the symmetry energy $S_0$, its slope $L_0$ and the incompressibility $K_0$. 

The symmetry energy $S_0$ is defined as the difference between E/A of symmetric and pure neutron matter
\begin{equation}
S_0=\frac{E}{A}(\rho_0,I=0)-\frac{E}{A}(\rho_0,I=1).
\end{equation}
$S_0(\rho)$ can be expanded about E/A with the second-order term being related to the \textit{asymmetry coefficient} $a_{sym}$  in the semi-empirical mass formula
\begin{equation}
a_{sym}=\frac{1}{2}\frac{\partial^2(E/A)}{\partial I^2}\Huge{|}_{I=0}.
\end{equation}
The slope of the symmetry energy, $L_0$ is 
\begin{equation}
L_0=3\rho_0\left(\frac{\partial{S}}{\partial\rho}\right)\Huge{|}_{\rho=\rho_0},
\end{equation}
and the incompressibility is calculated as
\begin{equation}
K_0=9\rho_0^2\frac{\partial^2(E/A)}{\partial\rho^2}\Huge{|}_{\rho=\rho_0}.
\end{equation}

\subsection{Observables for finite nuclei}
The requirement on input data for adjustment of the parameters of the QMC$\pi$-II is, as usual, to be known from experiment with high accuracy and least affected by correlations beyond mean-field. The first obvious choice is the ground state binding energy $BE$ which is directly available from solution of the mean field equations and can be readily extracted from highly precise measurements of atomic masses. 

The second choice relates to the density distribution of protons in the nucleus. Elastic electron scattering and optical methods provide information on charge density distribution in a nucleus and its mean-square charge radius, $\langle R^2_{ch}\rangle$. The model calculation provides mean-square radius of the proton distribution, $\langle R^2_p\rangle $, assuming the protons are point-like particles without internal structure. The two quantities are reasonably well related as \cite{Kort2010}
\begin{equation}
\langle R_{ch}^2 \rangle = \langle R_p^2\rangle+\langle r_p^2\rangle +\frac{N}{Z}\langle r_n^2\rangle ,
\label{eq:rch}
\end{equation}
with the free proton and neutron charge radii taken as $\langle r_p^2\rangle=0.7071$ fm$^2$  and $\langle r_n^2\rangle =-0.1161$ fm$^2$~\cite{Olive2014}. Note that the standard relation in Eq.~(\ref{eq:rch}) is valid for spherical nuclei. In Ref.~\cite{Bender2003}, additional term appears in Eq.~(\ref{eq:rch} )for charge radii in deformed nuclei.

Finally, because we work in the HF+BCS framework, data is needed to constrain parameters of the pairing EDF (\ref{eq:pair}), added to the QMC EDF. The pairing gap, a measure of nuclear pairing correlations, is a quantity that can be extracted from experimental  odd-even staggering in binding energies,  e.g. the neutron gap \cite{Klup2009}
\begin{equation}
\Delta^{(5)}_n = -\frac{1}{8}E(Z,N+2)+\frac{1}{2}E(Z,N+1)-\frac{3}{4}E(Z,N)+\frac{1}{2}E(Z,N-1)-\frac{1}{8}E(Z,N-2)
\end{equation}
and an equivalent expression for the proton gap. However, it is complicated to calculate the gaps in mean field models \cite{Bender2000,Klup2009}, in particular in a model including only even-even nuclei, and is not applicable if some of the nuclei is deformed. 

We therefore adopt as a measure of pairing correlations the average spectral gap  (for details see Ref.~\cite{Bender2000})
\begin{equation}
\bar{\Delta}_q=\frac{\Sigma_{\alpha\in q}u_\alpha v_\alpha \Delta_\alpha}{\Sigma_{\alpha\in q}u_\alpha v_\alpha} \, , 
\label{eq:delta}
\end{equation}
where $v_\alpha$, $u_\alpha=\sqrt{1-v_\alpha^2}$ are the occupation amplitudes and $\Delta_\alpha$ is the state-dependent single-particle pairing gap \cite{Schmid1987}.

As noted in \cite{Klup2009}, $\bar{\Delta}_q$ and $\Delta^{(5)}_{n,p}$ are reasonably well related in mid-shell regions but exhibit different behaviour in the vicinity of (semi)magic nuclei which may introduce a larger difference between experiment and model predictions.

\subsection{Parameter constraints}
The first stage of the fit required that the parameters of QMC$\pi$-II EDF satisfy the NMPs within their uncertainties.  The following limits were imposed on $\rho_0$, $E_0$,  $a_{sym}$, and $K_0$:
\begin{equation*}
\begin{split}
0.15\le \rho_0 \le 0.17 \quad (\text{fm}^{-3}), \\
-17\le E_0 \le -15  \quad (\text{MeV}),\\
29\le a_{sym} \le 33  \quad (\text{MeV}),\\
270\le K_0 \le 300  \quad (\text{MeV}).
\end{split}
\label{nmbounds}
\end{equation*}
In addition, the $\sigma$ meson mass $m_\sigma$ was also constrained to vary from 450 to 750 MeV but no constraints were applied on the other parameters. The search for all combinations of the QMC$\pi$-II parameters, satisfying these constraints resulted in a large number of parameter sets with errors too large to be meaningfully used in modeling finite nuclei.

\begin{center}
	\begin{figure}[h]
		\centering
	\includegraphics[angle=0,width=1.0\textwidth]{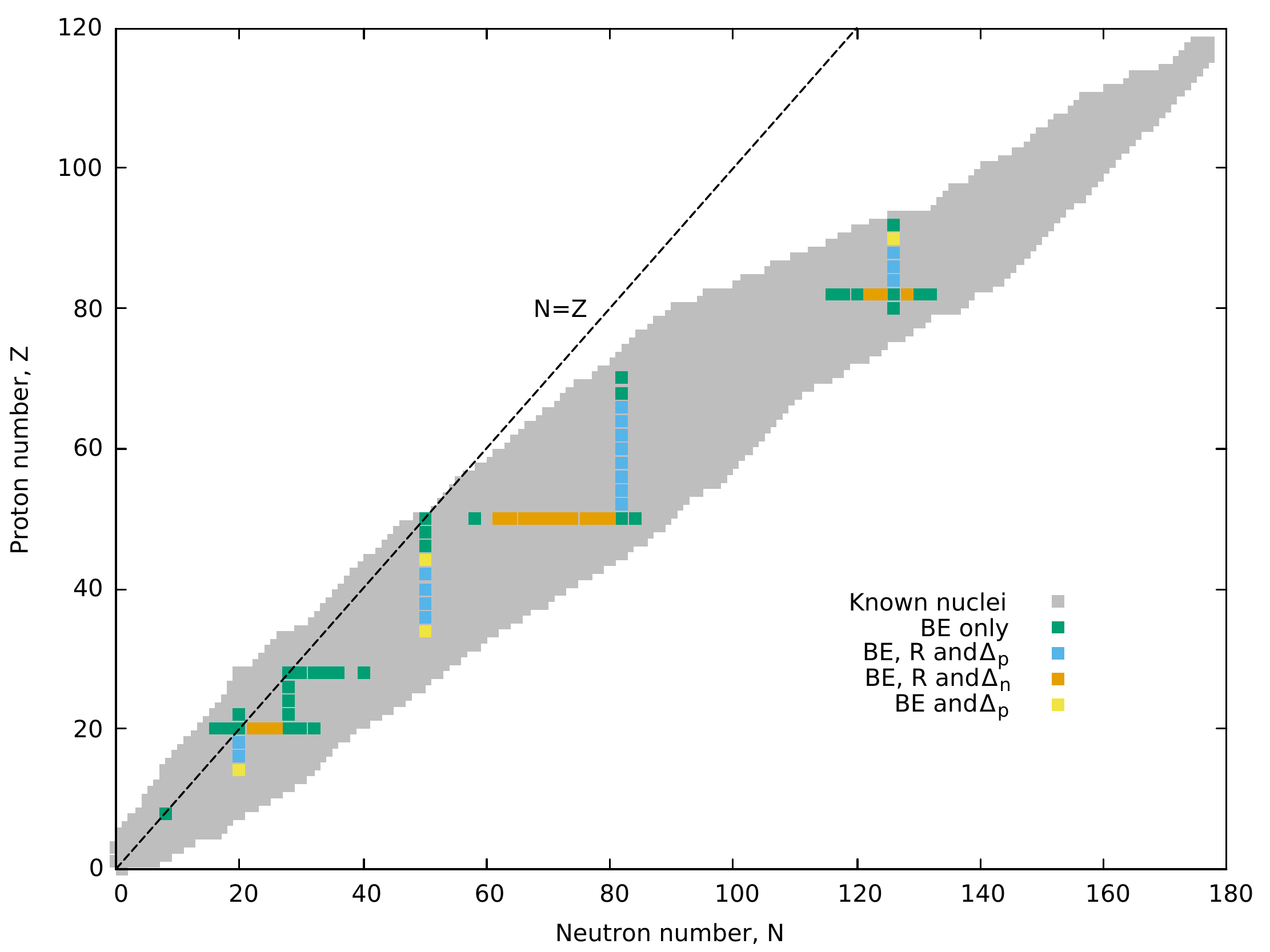}
	\caption{Doubly magic and semi-magic nuclei included in the fit. The nuclear observables and number of data points per a nucleus entering the fitting procedure are indicated. For more explanation see text.}
	\label{data}
	\end{figure}
\end{center}

To further narrow down the search for the best parameter set, a fit to selected experimental data was carried out. Binding energies $BE$, root-mean-square $rms$ charge radii $R_{ch}$ and proton and neutron pairing gaps, $\Delta_{p,n}$, calculated using Eq. (\ref{eq:delta}) for seventy magic and doubly-magic nuclei with $Z = 8, 20, 28, 50, 82$ and $N=126$ were included in the fit. This same set of even-even nuclei was used by Klupfel et al., \cite{Klup2009} to fit parameters for the Skyrme EDF with the exceptions of some updated values taken \cite{Wang2017} for binding energies and \cite{Angeli2013} for $rms$ charge radii. Figure~\ref{data} shows the distribution of the data across the nuclear chart, consisting of a total of 163 data points. 

\subsection{Parameter optimisation}
\label{pounders}
The search algorithm POUNDeRS, which stands for Parameter Optimization Using No Derivatives for 
the sum of squares~\cite{Kort2010} was used for the fitting procedure. POUNDeRS is a part of the Toolkit for Advanced Optimisation which is made available by the Portable, Extensible Toolkit for Scientific Computation (PETSc)~\cite{petsc-user-ref,petsc-efficient,tao-user-ref}. The algorithm has proven to be efficient in optimising nuclear energy density functionals of the Skyrme type \cite{Kort2010,Kort2012,Kort2013}. The main advantage of POUNDERS over other optimisation procedures is that it employs derivative-free algorithm, which is highly efficient in terms of the speed, accuracy and reliability of results~\cite{Wild2015}. Starting from initial values of the parameters, the total sum of squares of the deviations from experiment or the chi-squared value, $\chi^2$ is minimized. In this work, the objective function $F(\mathbf{\hat{x}})$ for minimisation was chosen to be dimensionless,
\begin{equation}
F(\mathbf{\hat{x}}) = \sum_i^{n}\sum_j^{p} \left(\frac{s_{ij}-\bar{s}_{ij}}{w_j}\right)^2,
\label{objfxn}
\end{equation}
where $n$ is the total number of nuclei, $p$ is the total number of observables and $s_{ij}$ and $\bar{s}_{ij}$ are the experimental and fitted values, respectively, for each nucleus $i$, and each observable $j$. $w_j$ stands for the \textit{effective} error for each observable, set in this fit to be  $w_{BE}=1$ MeV, $w_{R_{ch}}=0.02$ fm and $w_{\Delta_{p,n}}=0.12$ MeV for all nuclei without weighting. These errors, much higher than the errors reported by the experimenters, take into account a realistic estimate of accuracy of the model as well of the fitting procedure.

Following Kortelainen et. al.~\cite{Kort2010}, the covariance matrix was approximated as
\begin{equation}
\text{Cov}(\mathbf{\hat{x}}) \approx s^2(\mathbf{\hat{x}})\left(\mathbf{J^T} \mathbf{J}\right)^{-1},
\end{equation}
where $J$ is the Jacobian matrix with derivatives computed using finite differences and $s^2=\chi^2/(n-p)$. The objective function was evaluated at \{$\mathbf{\hat{x}}\pm\eta e_j$\}, with $\eta$ set to $10^{-3}$ and  $e_j$ being the scale used for each parameter during the search. 

The square root of the diagonal terms of the covariance matrix gives the \textit{standard deviation} $\sigma$ for each parameter and the off-diagonal terms give the correlation coefficient between any two parameters $x_k$ and $x_l$
\begin{equation}
\text{Cor}(x_k,x_l) = \frac{\text{Cov}(x_k,x_l)}{\sqrt{\sigma_{x_k}^2\sigma_{x_l}^2}}.
\end{equation}
A \textit{residual} is defined as the difference between the experimental and theoretical  results, $s_{ij} - \bar{s}_{ij}$, and is used to evaluate the \textit{root-mean-square deviation} (RMSD) for each observable
\begin{equation}
\text{RMSD}(j) = \sqrt{\frac{1}{n}\sum_i^{n} (s_{ij} - \bar{s}_{ij})^2}. 
\label{rmsd}
\end{equation}
The \textit{percentage deviation} from experiment is $100*\left(\frac{s_{ij} - \bar{s}_{ij}}{\bar{s}_{ij}}\right)$.

\section{Results and discussion}
\label{results}
In this section the results of the optimisation are presented. The final parameter set has been used to calculate all observables for the 70 nuclei included in the fit (Sec.~\ref{70}) and further applied to calculate binding energies and charge radii for even-even nuclei with experimentally known masses with $Z\ge8$ (Sec.~\ref{ext}). Predictions of the model for other observables not included in the fit are discussed (Sec.~\ref{obsnot}), along with calculations of ground state binding energies of superheavy even-even nuclei (Sec.~\ref{SHE}).

\subsection{The parameters}
\label{par}
Table~\ref{pars} summarizes the best fit QMC$\pi$-II parameter set with their confidence intervals and standard deviations. The NMPs corresponding to this set are $\rho_0 = 0.15 \pm 0.01$ fm$^{-2}$, $E_0 = -15.78 \pm 0.02$ MeV, $a_{sym} = 30.6 \pm 0.3$ MeV and $K_0=270\pm2$ MeV.
\begin{table}[ht]
	\centering
	\caption{Confidence intervals (CI)  and standard deviation $\sigma$ of the final QMC$\pi$-II parameter set. The proton and neutron pairing strengths are included for completeness.}
\begin{ruledtabular}
	\begin{tabular} {c c c c}
		Parameter &  Value  & 95\% CI &  $\sigma$ \\	
		\hline
		$G_\sigma$  [fm$^2$]&9.05&[9.02, 9.08]&0.01\\
		$G_\omega$  [fm$^2$]&5.29&[5.27, 5.30]&0.01\\
		$G_\rho$  [fm$^2$]&4.71&[4.64, 4.78]&0.04\\
		$m_\sigma$ [MeV]&	 495 &[491, 498]&2\\
		$\lambda_3$&0.049&[0.048, 0.050]&0.001\\ 
 		$V_p^{pair}$ [MeV]&288&[280, 295]&4\\
		$V_n^{pair}$ [MeV]&275&[261, 290]&7\\             
	\end{tabular}
	\label{pars}
\end{ruledtabular}
\end{table}
One important feature of  QMC$\pi$-II is the smaller value for the incompressibility $K_0=270\pm2$ MeV, which tended to be somewhat high in the previous models, QMC-I~\cite{Stone2016} and QMC$\pi$-I~\cite{Guichon2018}, where the values were 340 MeV and 319 MeV, respectively. Recent re-analysis of data from giant monopole resonances provided a range for $K_0$ values from 250 MeV to 315 MeV~\cite{Stone2014}. Another significant result is that the slope of the symmetry energy $L_0$, not included in the fits but calculated afterwards with the final parameter set,
is $L_0=70$ MeV, which is now closer to the value expected from various analyses. Ref.~\cite{Li2013} summarises 28 available results from various terrestrial measurements and astrophysical observations for the symmetry energy and its slope, with average values of 31.6 MeV and 58.9 MeV, respectively. The finite-range droplet model combined with folded Yukawa microscopic part (FRDM)~\cite{Moller2016}, for example, has values $a_{sym}$=32.5$\pm$ 0.5  MeV and $L_0$=70$\pm$ 15 MeV. Recently, by studying the radioactivity of 19 proton emitters having large isospin asymmetry, $L_0$ is constrained to have a value of 51.8 $\pm$ 7.2 MeV~\cite{Wan2016}. The previous version  of the QMC model, QMC-I~\cite{Stone2016} and QMC$\pi$-I \cite{Guichon2018,Stone2017}, gave $L_0$=23$\pm$ 4 MeV and $L_0$=17$\pm$1 MeV, respectively.  

Obviously, all QMC EDFs have correlated parameters and they vary accordingly in the optimisation procedure. For the QMC$\pi$-II final parameter set, these correlations are computed as discussed in section~\ref{pounders}. 
Table~\ref{cisd} shows the correlation between any two parameters of the EDF. A positive (negative) correlation means that parameters are directly (inversely) proportional to each other and a value of 1.0 corresponds to 100\% correlation.

\begin{table}[h]
	\centering
	\caption{Correlation between QMC$\pi$-II parameters computed as discussed in section \ref{pounders}}
\begin{ruledtabular}
	\begin{tabular} {c| c c c c c c c}
		&$G_\sigma$&$G_\omega$&$G_\rho$&$m_\sigma$&$V_n^{pair}$&$V_n^{pair}$&$\lambda_3$\\	
		\hline
	$G_\sigma$	&1.00&&&&&&\\
	$G_\omega$	&0.91	&1.00&&&&&\\
	$G_\rho$	&-0.22	&0.18	&1.00&&&&\\
	$m_\sigma$	&0.18	&-0.07	&-0.72 &1.00&&&\\
	$V_p^{pair}$		&0.02	&0.14	&0.30  &-0.22  &1.00&&\\
	$V_n^{pair}$		&-0.01	&0.02	&0.05  &-0.06  &0.03  &1.00&\\
	$\lambda_3$	&-0.22	&0.13	&0.97  &-0.79  &0.30 &0.05 &1.00\\
	\end{tabular}
	\label{cisd}
\end{ruledtabular}
\end{table}
Strong positive correlation is seen between the effective couplings $G_\sigma$ and $G_\omega$ as well as between $G_\rho$ and the self-coupling $\lambda_3$ parameter. On the other hand, the $\sigma$ meson mass is inversely dependent on both $G_\rho$ and $\lambda_3$. The introduction of the new $\lambda_3$ parameter in the QMC$\pi$-II model led to a decrease in the three coupling parameters and the value of $m_\sigma$ compared to the QMC-I parameters~\cite{Stone2016}, while it effectively tuned down the incompressibility, $K_0$. For the parameters of the pairing EDF, the proton pairing strength has 30\% correlation to both $G_\rho$ and $\lambda_3$ while neutron pairing strength showed little correlation with the couplings and $\lambda_3$ parameters. This means that changes in the coupling parameters and thus NMP values do not significantly affect the pairing strengths.

\subsection{Nuclei included in the fit}
\label{70}
Table~\ref{rmsfit} shows a summary of \textit{percent} deviations of the observables for the seventy finite nuclei included in the fit in comparison with the previous QMC-I~\cite{Stone2016} and QMC$\pi$-I~\cite{Stone2017,Guichon2018} and results from Skyrme force SV-min~\cite{Klup2009}. 
It should be noted that QMC-I, QMC$\pi$-I and SV-min fits included data on diffraction radii and surface thickness which were not included in the current fit for QMC$\pi$-II.
\begin{table}[ht]
\begin{ruledtabular}
	\centering
	\caption{Percent deviations of observables of nuclei included in the QMC$\pi$-II  fit. QMC-I \cite{Stone2016} and QMC$\pi$-I \cite{Stone2017,Guichon2018} are results from previous versions of the QMC model. The results from Skyrme SV-min \cite{Klup2009} is added for comparison.}
	\begin{tabular} {l c c c c}
		Data& QMC$\pi$-II & QMC$\pi$-I & QMC-I & SV-min\\
		\hline
\quad	Binding energy & 0.34 &0.46&0.36 & 0.24\\
\quad	\textit{rms} charge radius & 0.59 &0.48& 0.71 & 0.52\\
\quad	Proton pairing gap & 25.7 &15.3& 25.3 & 15.5\\
\quad	Neutron pairing gap & 16.1 &24.0& 57.6 & 17.6\\
	\end{tabular}
	\label{rmsfit}
\end{ruledtabular}
\end{table}
The QMC$\pi$-II model, with inclusion of the $\sigma$ self-interaction as well as of the single-pion exchange, together with the current fitting procedure, showed some notable improvements in the predictions for the nuclear observables compared to previous QMC versions. Significantly, the neutron pairing gap \textit{percentage} deviation is now lower than that of QMC-I and QMC$\pi$-I. Binding energies are also improved in QMC$\pi$-II but charge radii and proton pairing gaps were better in QMC$\pi$-I.

\begin{center}
	\begin{figure*}[ht]
		\centering
		\includegraphics[angle=0,width=1.0\linewidth]{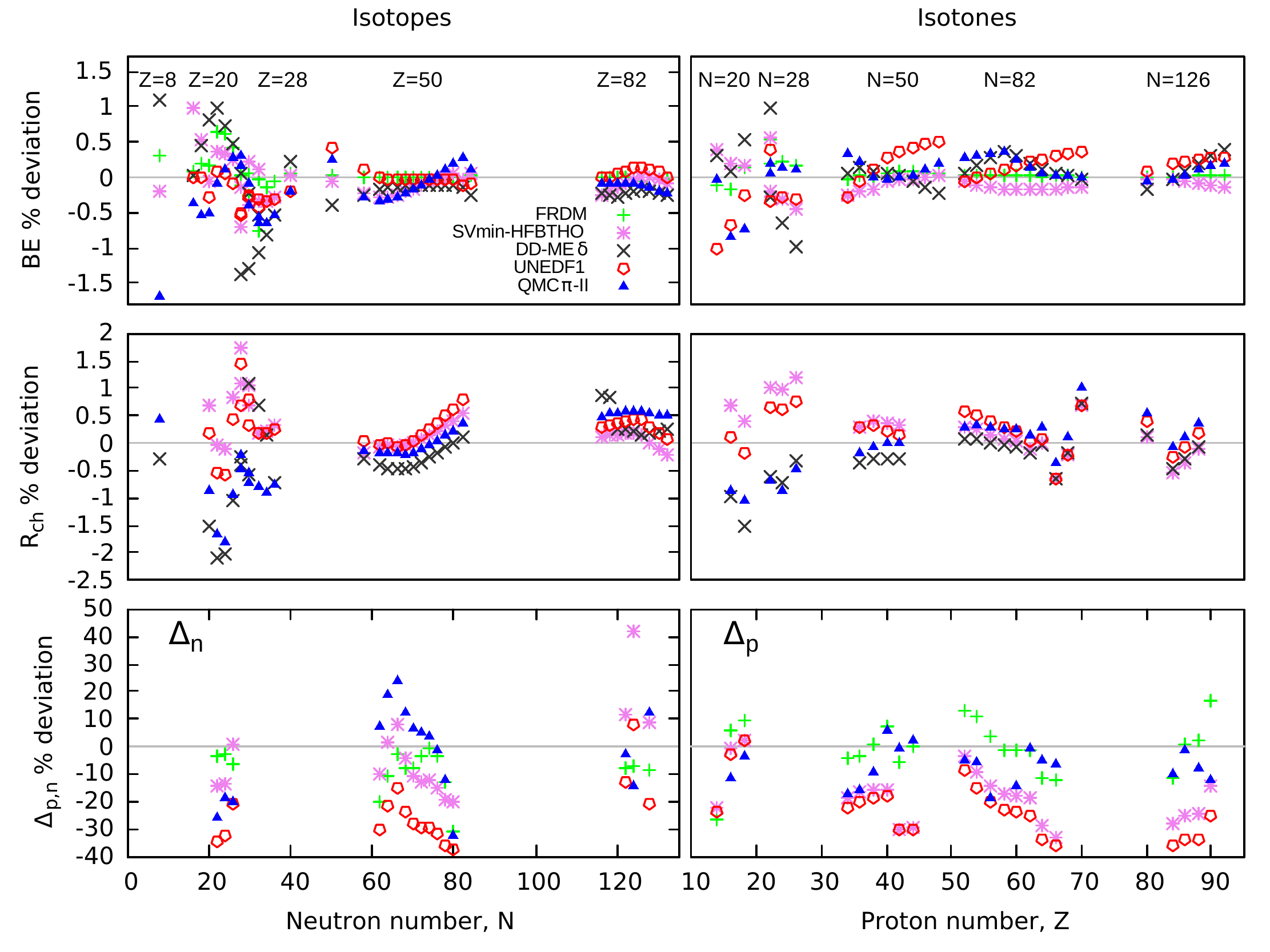}
		\caption{Percent deviation from experiment for binding energies $BE$, \textit{rms} charge radii $R_{ch}$, and pairing gaps $\Delta_{p,n}$ for the 70 nuclei in the fit. Added for comparison are the Skyrme type force SV-min~\cite{Klup2009} and UNEDF1~\cite{Kort2012}, a covariant energy density DD-ME$\delta$~\cite{Afa2016} and the finite-range droplet model (FRDM)~\cite{Moller2016}. The plot legend is located in the top left panel.}
		\label{dev}
	\end{figure*}
\end{center}

 Figure~\ref{dev} illustrates the \textit{percent} deviation from experiment for $BE$, $R_{ch}$, and $\Delta_{p,n}$ for the 70 nuclei, including Z=20, 28, 50 and 82 isotopes and N=20, 28, 50,82 and 126 isotones. It can be seen that the QMC$\pi$-II results follow almost the same trend as the other models, having relatively higher deviations for lighter nuclei $Z,N<28$ for binding energies, $BE$, and charge radii, $R_{ch}$. The $BE$ absolute deviations are up to 1.7\%, with the highest value for $^{16}$O, while $rms$ charge radii \textit{absolute} deviations are up to 1.8\%. Pairing gaps deviations are typically within 30\%.
\subsection{Extended fit to binding energies and charge radii across the nuclear chart. }
\label{ext}
\begin{center}
	\begin{figure}[ht]
		\centering
		\begin{tabular}{c}
			\includegraphics[angle=0,width=0.9\textwidth]{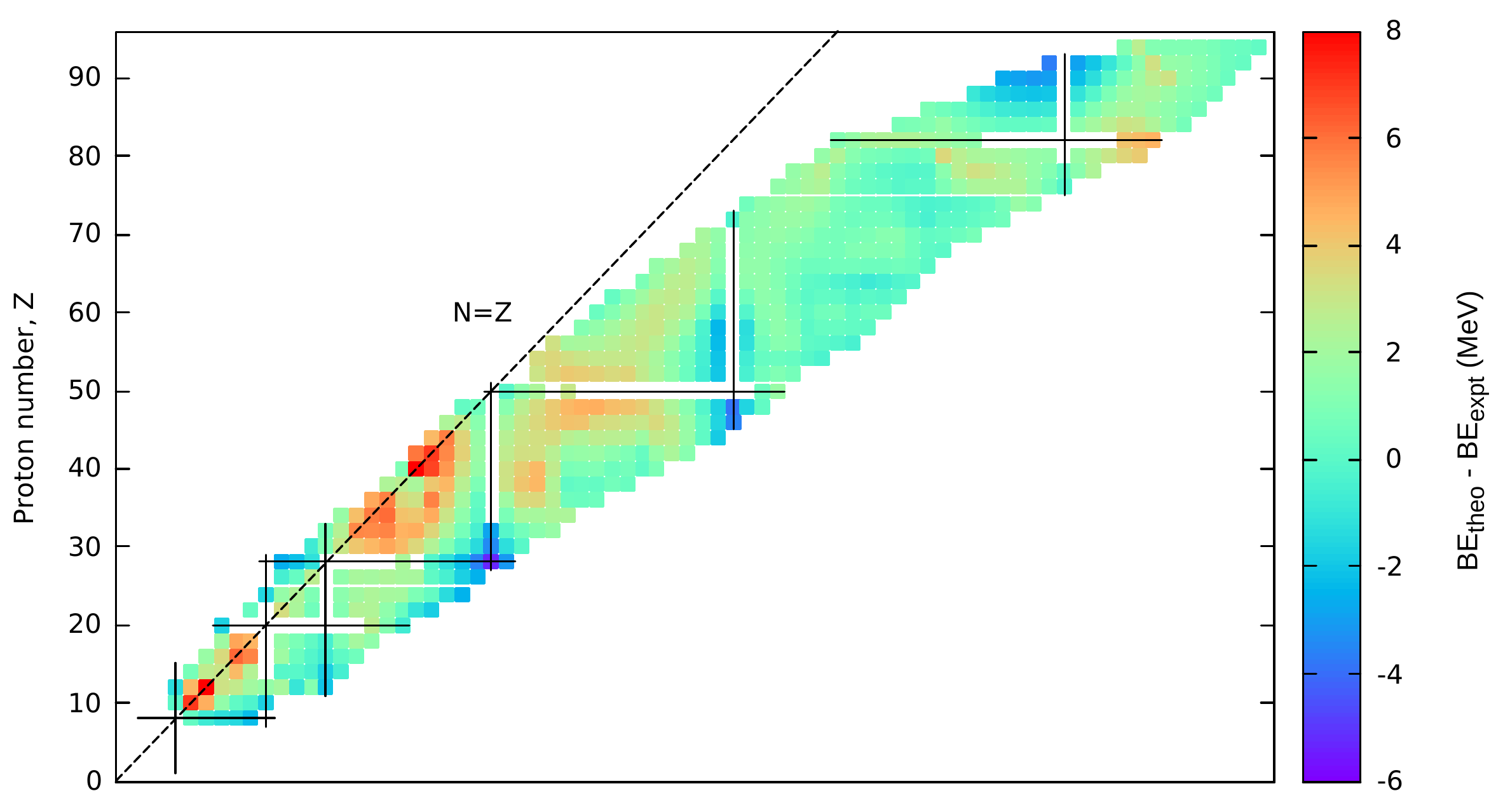}\\
			\includegraphics[angle=0,width=0.9\textwidth]{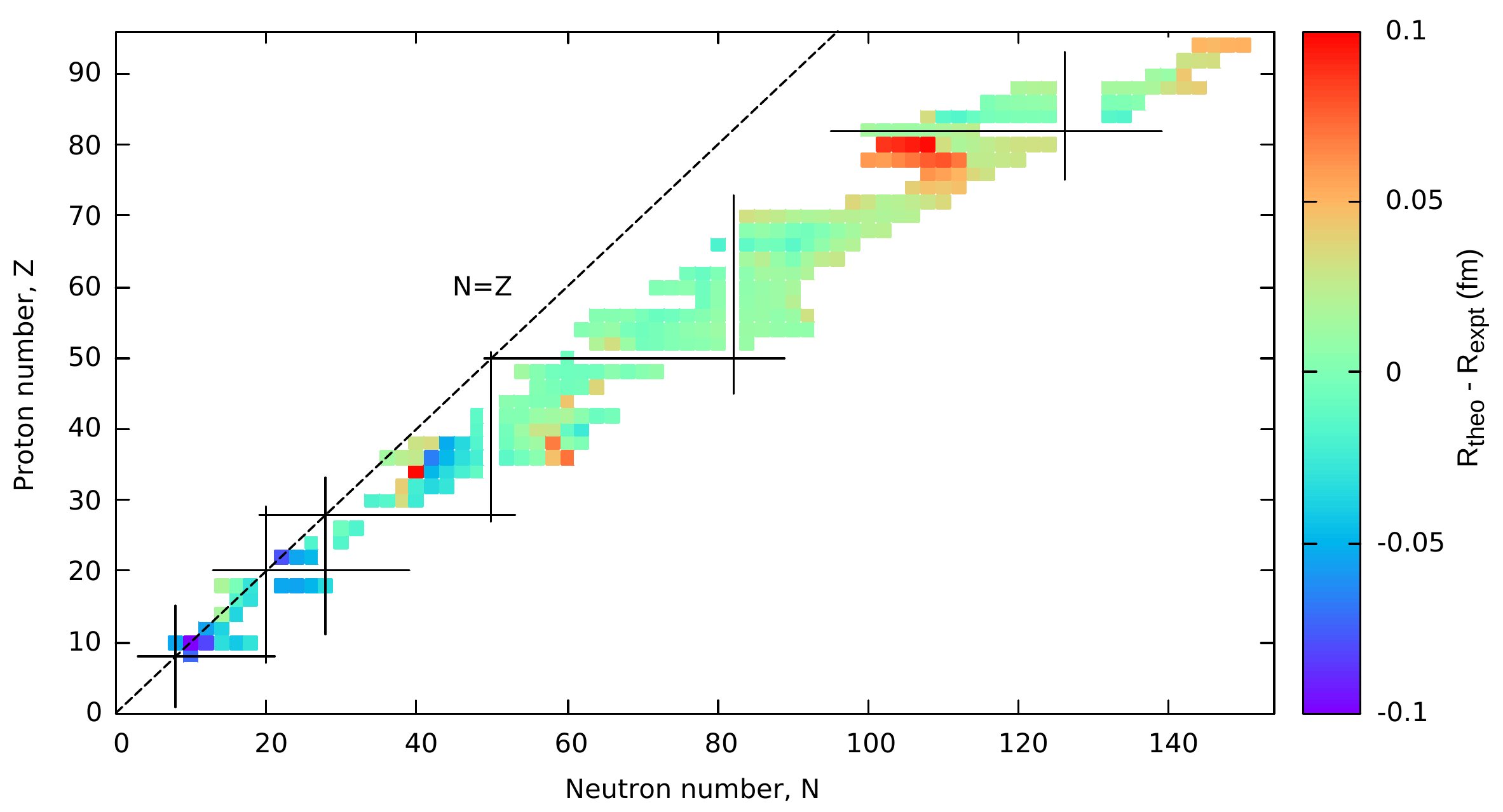}\\
		\end{tabular}
		\caption{Binding energy and \textit{rms} charge radii residuals for 669 even-even nuclei with $Z < 96$. Atomic mass data used to compute the binding energy residuals are taken from~\cite{Wang2017} and $rms$ charge radii data are from~\cite{Angeli2013}. Nuclei with magic numbers are indicated by solid lines.}
		\label{resid}
	\end{figure}
\end{center}

Figure~\ref{resid} illustrates the performance of QMC$\pi$-II for a set of 669 even-even nuclei with known masses, 286 of them have known $rms$ charge radii. This set excludes the seventy nuclei included in the QMC$\pi$-II fit which was already discussed in Sec. \ref{70}. The binding energy residuals vary within around -6 to 8 MeV and the charge radius residuals are within $\pm0.1$ fm. These are essentially the same ranges found for SV-min, DD-ME$\delta$ and UNEDF1 but with a different distribution of residual values across the nuclear chart. 
The QMC$\pi$-II RMSD for all nuclei included in the plot is 2.23 MeV for masses and 0.030 fm for radii.

The current QMC$\pi$-II parameter set predicts overbinding in most of the mirror nuclei and the residuals are relatively higher for these nuclei compared with the other models. $N=Z$ nuclei are known to exhibit  Wigner effect~\cite{Qi2015} that must be accounted for in the binding energy.  Furthermore, the QMC$\pi$-II parameter set predicts mostly underbinding on the neutron-rich side, as shown in Figure~\ref{resid}. There are also relatively larger errors around the magic isotones N = 50 and 82 and in the uranium region.

For charge radii in Figure~\ref{resid}, note that they have been calculated using the standard relation in Eq.~(\ref{eq:rch}). QMC$\pi$-II residuals for $R_{ch}$ are relatively higher near the $Z=82$ shell closure, specifically in the mercury ($Z=80$) region. Neutron-deficient lead isotopes are mostly spherical from laser spectroscopy experiments but mercury and platinum isotopes with neutron number around $N=104$ show deformations when compared with droplet model calculations~\cite{DeWitte2007,Neugart2017}.

To compare the performance of QMC$\pi$-II with other models, binding energy and \textit{rms} charge radii residuals were also computed for exacly the same set of nuclei included in Fig. \ref{resid}. Table \ref{table:resid} presents the RMSD computed using Eq. (\ref{rmsd}). Overall, QMC$\pi$-II results appear to be more or less on par with other models. FRDM gives the best predictions for masses with an RMSD of 0.67 MeV but there is no available record of its predictions for charge radii. 
\begin{table}[ht]
	\begin{ruledtabular}
		\centering
		\caption{RMSD for binding energies and \textit{rms} charge radii for QMC$\pi$-II in comparison with Skyrme forces SV-min \cite{Klup2009}, UNEDF1 \cite{Kort2012}, covariant EDF with the DD-ME$\delta$ interaction \cite{Afa2016,frib}, and macroscopic-microscopic FRDM \cite{Moller2016}.}
		\begin{tabular} {l c c c c c}
			Data& QMC$\pi$-II & SV-min & UNEDF1 & DD-ME$\delta$ & FRDM\\
			\hline
			\quad	Binding energy (MeV)	& 2.23 &3.22 & 2.11& 2.35& 0.67\\
			\quad	\textit{rms} charge radius (fm)	& 0.03 &0.02 & 0.03 &0.04&- \\
		\end{tabular}
		\label{table:resid}
	\end{ruledtabular}
\end{table}
\subsection{Observables not included in the fit}
\label{obsnot}
\subsubsection{Two-nucleon separation energies} 
\begin{center}
\begin{figure}[h]
	\centering
	\includegraphics[angle=0,width=1.0\textwidth]{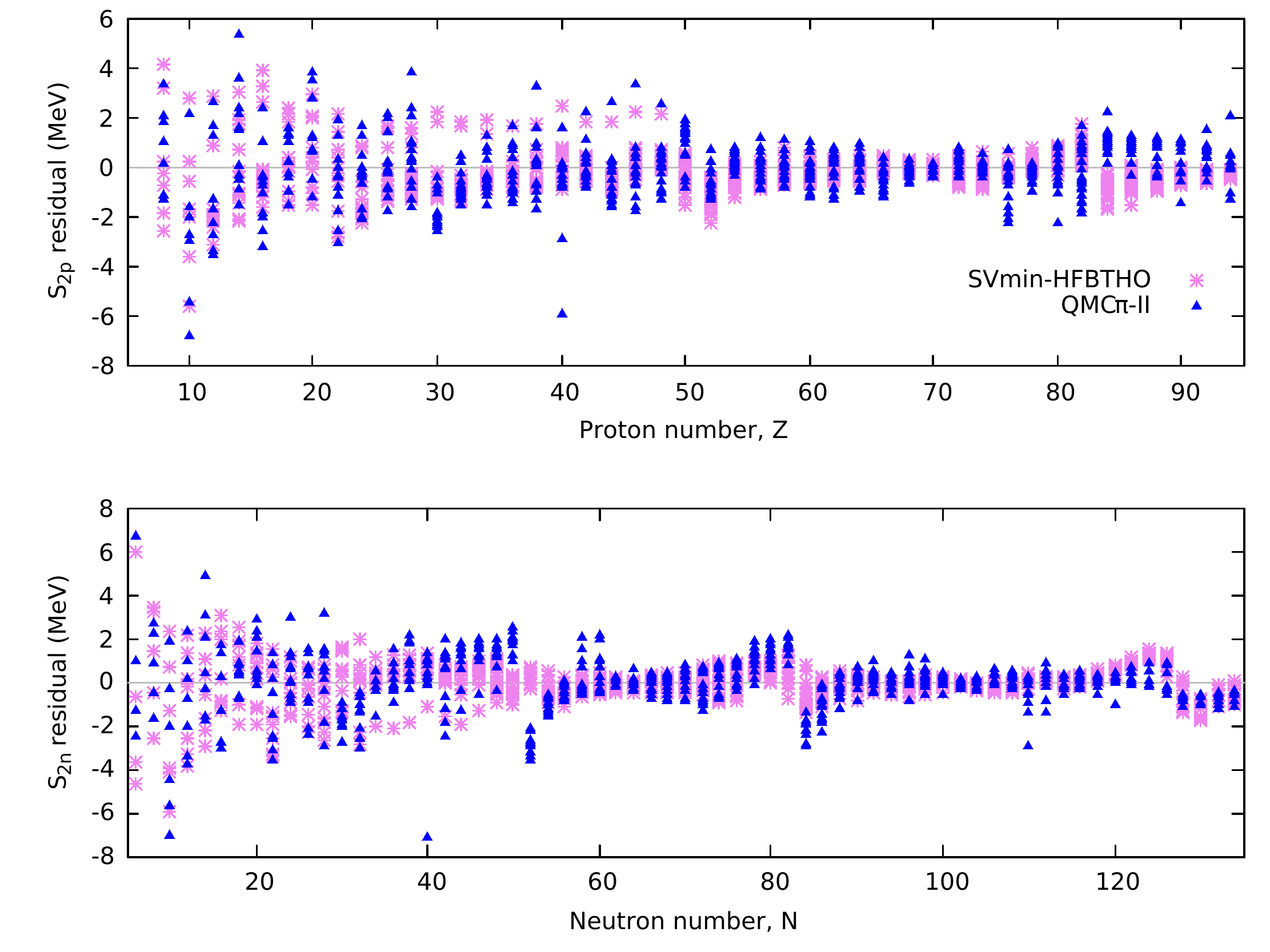}
	\caption{Two-nucleon separation energy residual for even-even isotopes and isotones for QMC$\pi$-II and SV-min. Experimental data used to compute the residuals are taken from 
		Ref.~\cite{Wang2017}. The plot legend is located in the top panel.}
	\label{sepens}
\end{figure}
\end{center}

Two-nucleon separation energies provide and important information about the existence and location of driplines.
Figure~\ref{sepens} shows the residuals for even-even nuclei with available data for two-proton ($S_{2p})$ and two-neutron ($S_{2n}$) separation energies, comparing results from QMC$\pi$-II and SV-min. Both models are within the same range of residuals.

\begin{center}
	\begin{figure}[ht]
        \centering
		\includegraphics[angle=0,width=1.0\textwidth]{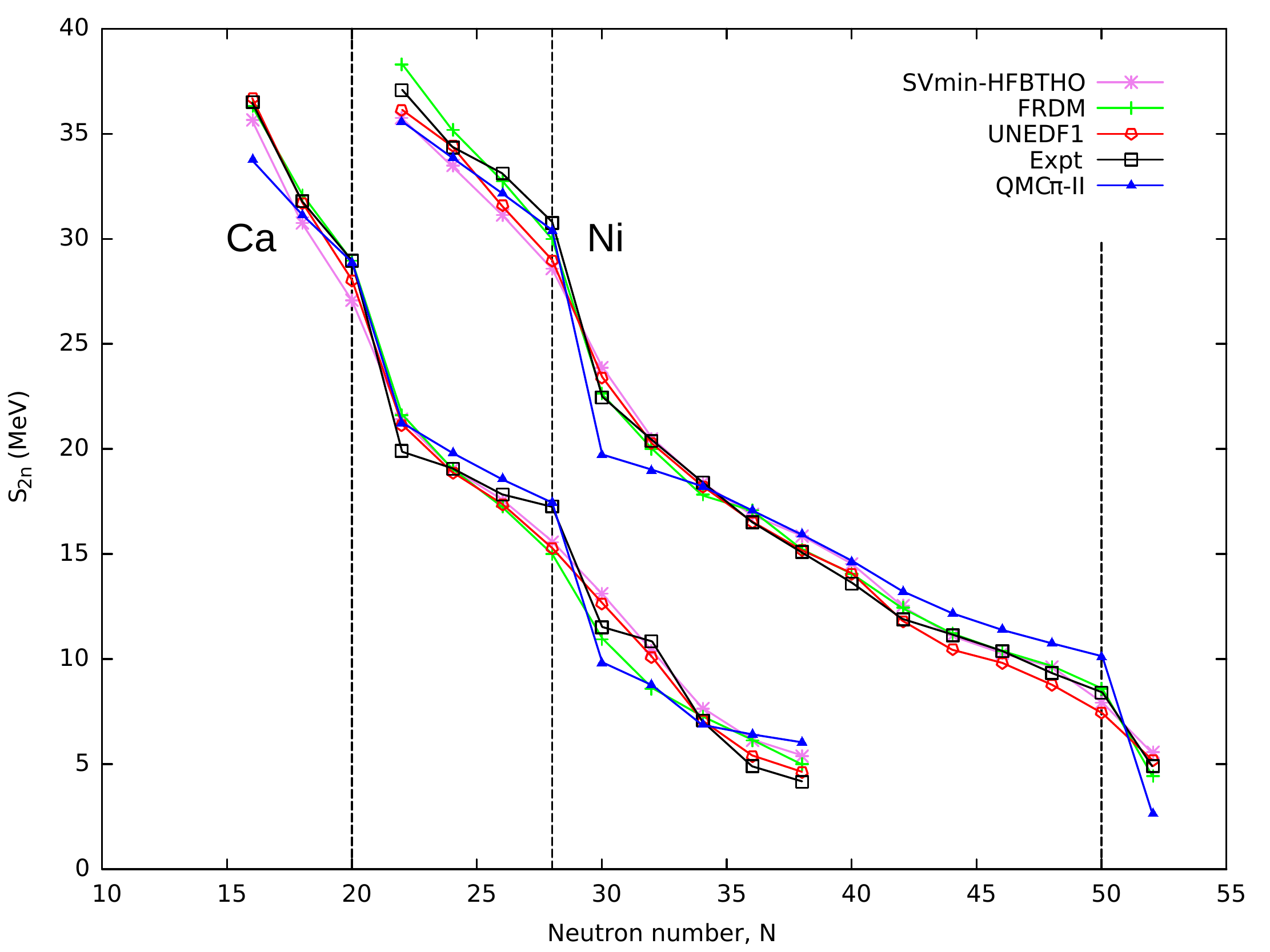}\\
		\caption{Two-neutron separation energies for calcium ($Z=20$) and nickel ($Z=28$) isotopes as a function of the neutron number. Shell closures are visible at $N=20$ for Ca, $N=28$ for both Ca and Ni and $N=50$ for Ni isotopes and are indicated by dashed lines. Also added for comparison are results for FRDM \cite{Moller2016}, SV-min~\cite{Klup2009} and UNEDF1~\cite{Kort2012}, taken from Ref.~\cite{frib} and experimental data from \cite{Wang2017}.}
		\label{S2nCaNi}
	\end{figure}
\end{center}

\begin{center}
	\begin{figure}[h]
        \centering
		\includegraphics[angle=0,width=1.0\textwidth]{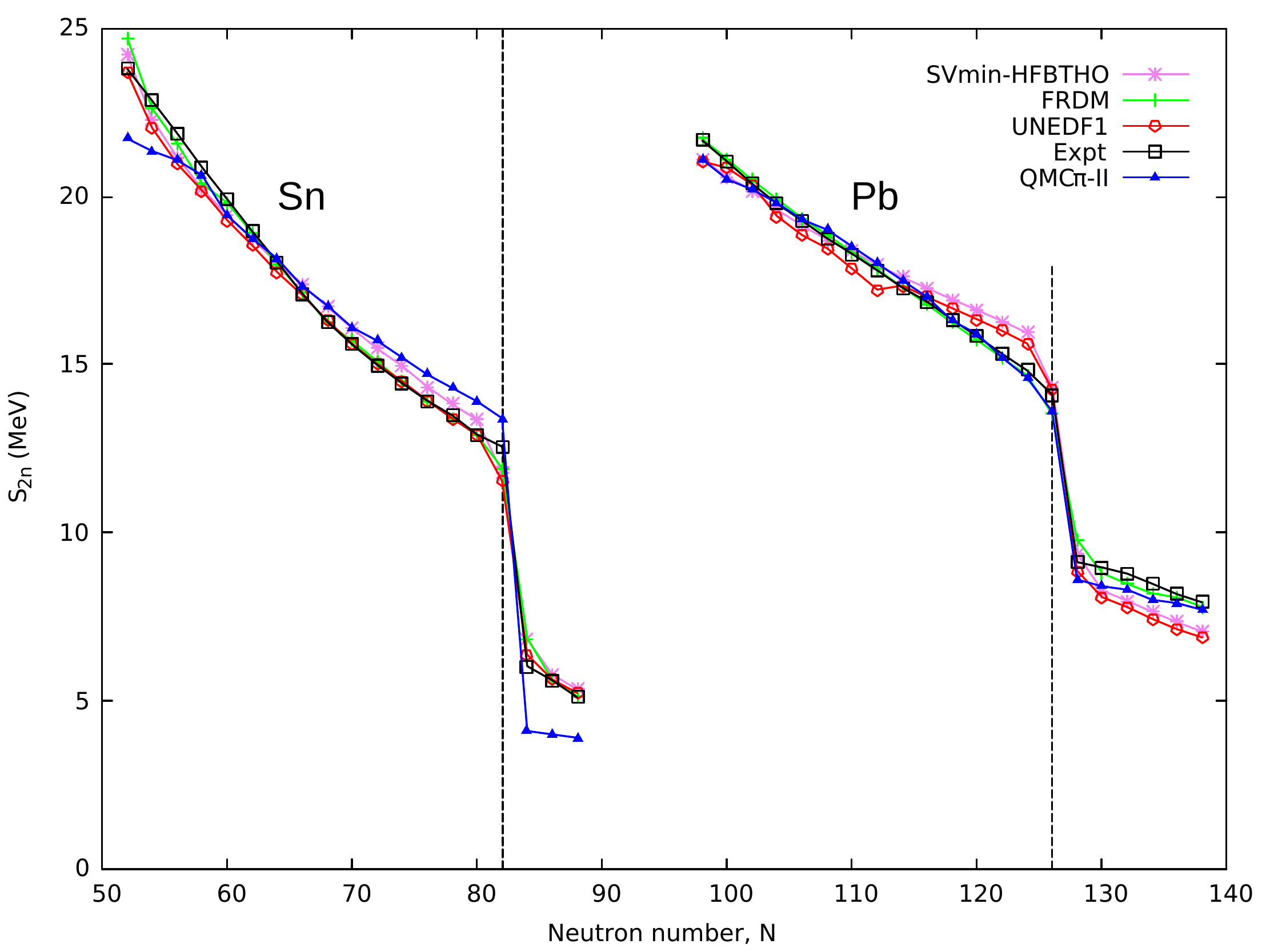}\\
		\caption{Same as in Fig. \ref{S2nCaNi} but for tin ($Z=50$) and lead ($Z=82$) isotopes. Shell closures are visible at $N=82$ for Sn and $N=126$ for Pb isotopes and are indicated by dashed lines.}
		\label{S2nSnPb}
	\end{figure}
\end{center}

In particular, two-neutron separation energies for the magic isotopes calcium, nickel, tin and lead are shown in Figures~\ref{S2nCaNi} and \ref{S2nSnPb}. Shell closures at $N=20$ for Ca, $N=28$ for the Ca and Ni isotopes and $N=50$ for Ni are visible through the sudden dip in the separation energies. The same is true for shell closures at $N=82$ and $N=126$ for the Sn and Pb isotopes, respectively.

Compared to other models in Fig.~\ref{S2nCaNi}, QMC$\pi$-II shows pronounced shell closure at $N=28$ for both calcium and nickel and at $N=50$ for nickel. The deviation from experiment is, however, relatively larger at $N=30$ and 32, as well as around $N=50$ and $N=82$. QMC$\pi$-II results around the $N=126$ closure, as shown in Figure~\ref{S2nSnPb} are in good agreement with experiment.

\subsubsection{Isotopic shifts in charge radii}
\begin{center}
	\begin{figure*}[ht]
		\centering
		\includegraphics[angle=0,width=1.0\textwidth]{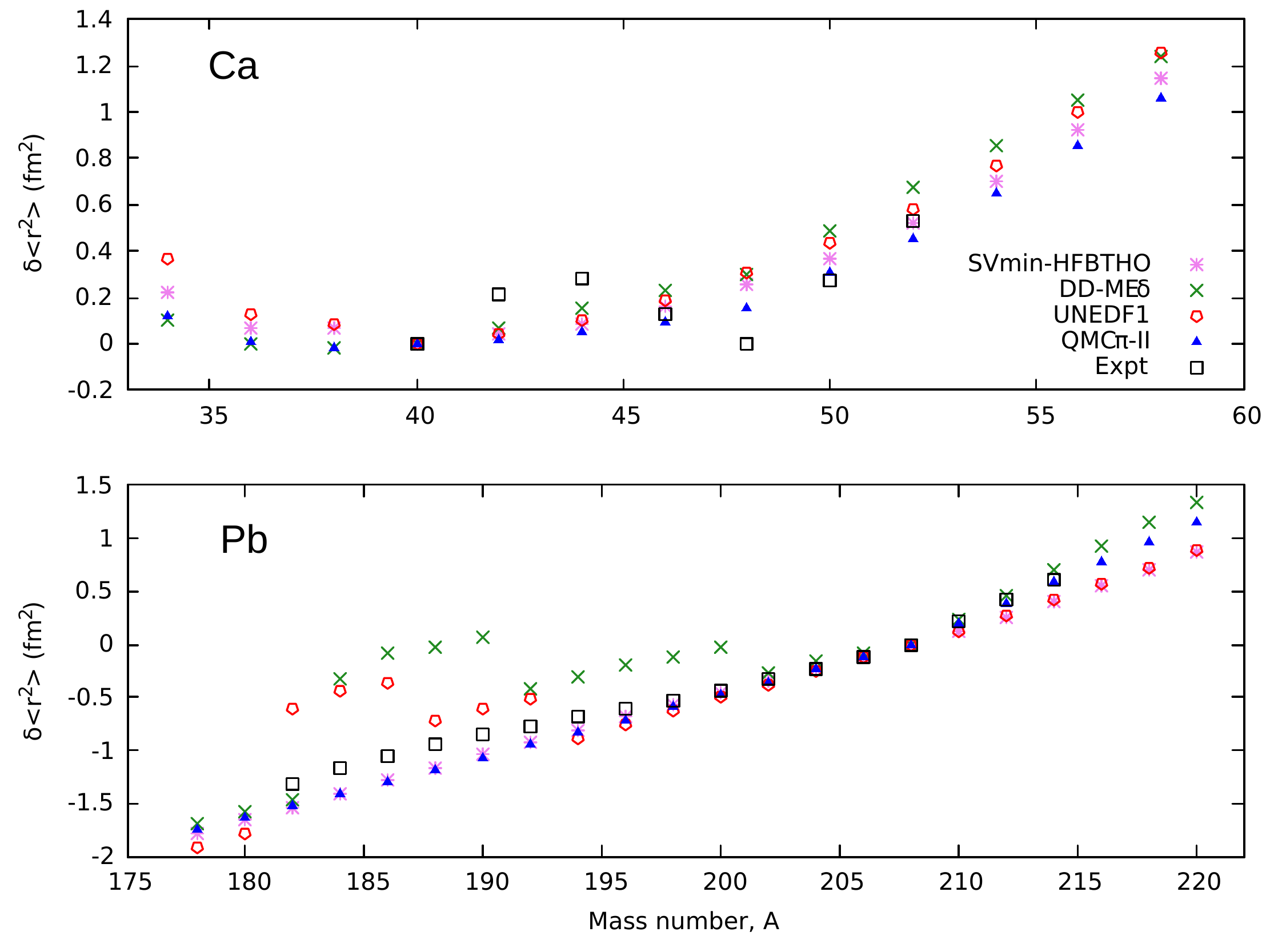}
		\caption{Isotopic shifts for calcium and lead isotopes with stable reference isotopes $^{40}$Ca and $^{208}$Pb, respectively, plotted against mass number $A$. Added for comparison are values for SV-min \cite{Klup2009}, DD-ME$\delta$ \cite{Afa2016} and UNEDF1 \cite{Kort2012}, all taken from Ref.~\cite{frib}. Experimental data are from \cite{Angeli2013} and errors are smaller than the symbols used in the plot. Plot legend is located in the top panel.}
		\label{radshift}
	\end{figure*}
\end{center}

 In an isotopic chain, charge radii evolution with mass number is characterised by the difference between mean-square charge radii of two isotopes of an element calculated using $\delta\langle r^2\rangle^{A',A} = \langle r^2\rangle ^{A'} -\langle r^2\rangle^A$. Here $\langle r^2\rangle$ is the mean-square charge radius for the isotope with nucleon number $A'$ and $A$ is the reference isotope. Figure~\ref{radshift} shows the shifts in radii for calcium and lead from their stable reference isotope $^{40}$Ca and $^{208}$Pb, respectively. Experimental data are taken from Ref.~\cite{Angeli2013} with updates for Ca isotopes from Ref.~\cite{Ruiz2016}.

The QMC$\pi$-II radius computations for both the Ca and Pb isotopes are done under the constraint that they should be spherical. For Ca isotopes, none of the models reproduce the the trend of the experimental data where the radius shift initially increases from $^{40}$Ca and then drops to almost zero at $^{48}$Ca. All models predicted a continued increase in the radius shift, starting from $^{40}$Ca. Good agreement is seen, however, for the recently measured shift to $^{52}$Ca with an experimental value of 0.531(5) fm$^2$. However, taking $^{48}$Ca as the stable reference isotope for isotopic shift to $^{52}$Ca, QMC$\pi$-II predicts a radius shift of 0.300 fm$^2$, while SV-min gives 0.262 fm$^2$, while the recent data yields 0.530(5) fm$^2$~\cite{Ruiz2016}. 
Certainly this behaviour will be studied further as the QMC model continues to be developed. For lead isotopes, QMC$\pi$-II and SV-min have almost the same behaviour for radius shifts in the neutron-deficient region, while UNEDF1 predicts oblate deformation for Pb isotopes around $A = 190$, which is caused by its low proton state gap at $Z=82$~\cite{Kort2012}. Radius shifts from DD-ME$\delta$ are also higher than those found experimentally for isotopes with $A<200$.  On the neutron-rich side of the Pb isotopes, QMC$\pi$-II agrees well with experiment, while SV-min and UNEDF1 give lower radii shifts as the mass number increases. The isotopic shift from $^{208}$Pb to $^{214}$Pb was considered in Ref.~\cite{Klup2009}, with different values for the effective mass. For QMC$\pi$-II, this Pb radius shift has a value of 0.589 fm$^2$, which is close to the measured value is 0.615$\pm$0.001 fm$^2$.

\subsubsection{Neutron skin thickness}
\begin{center}
	\begin{figure*}[h]
		\centering
		\includegraphics[angle=0,width=1.0\textwidth]{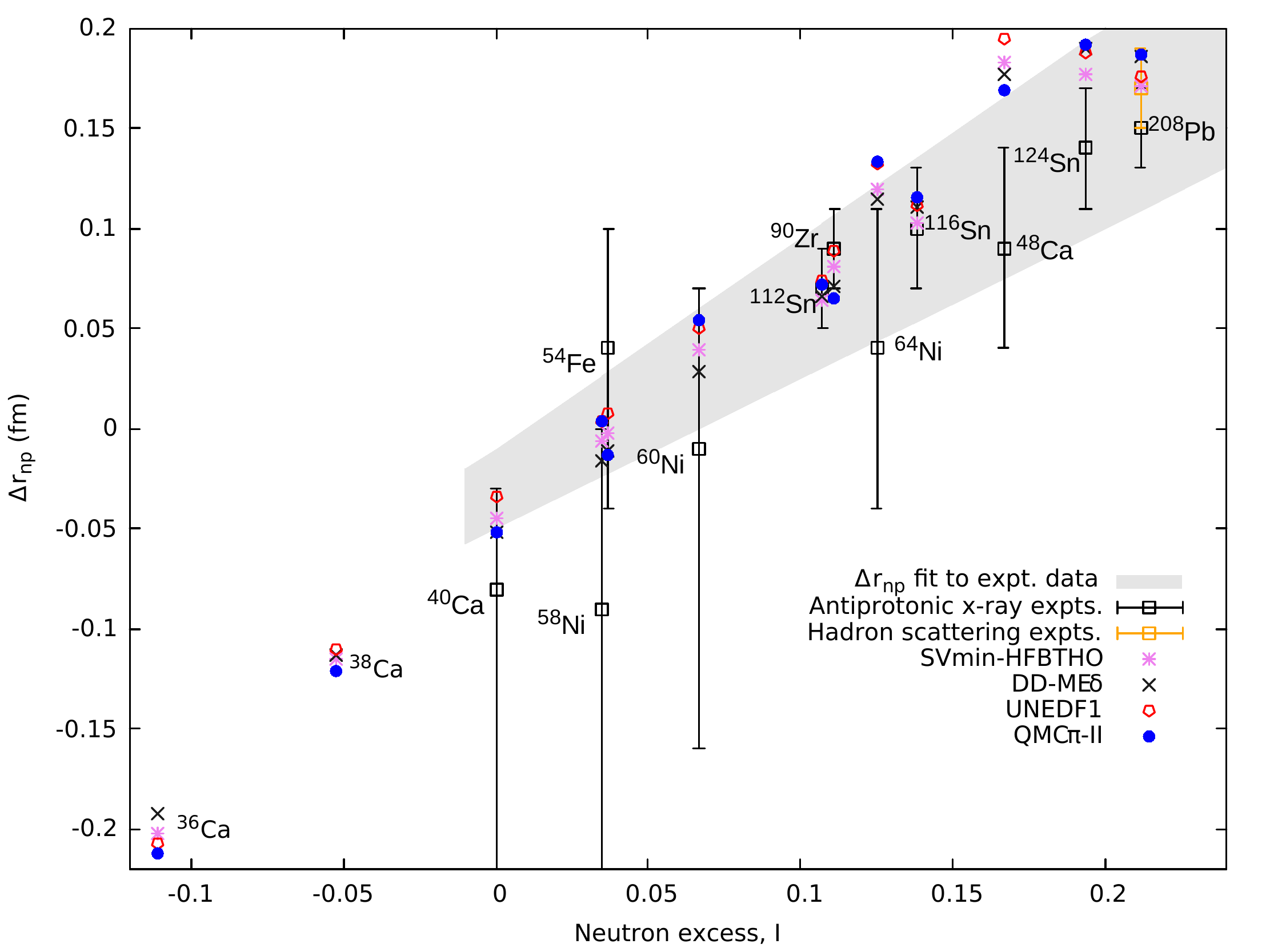}
		\caption{Skin thickness for nuclei included in the QMC$\pi$-II fit as a function of the neutron excess $I=(N-Z)/A$. Added for comparison are SV-min \cite{Klup2009} and DD-ME$\delta$ \cite{Afa2016} both taken from \cite{frib} and experimental data from \cite{Trz2005}.}
		\label{skin}
	\end{figure*}
\end{center}

Another observable relating to size, which is of considerable interest, is the neutron skin thickness, $\Delta r_{np}$, defined as the difference between the neutron and proton point radii. Neutron skin thickness has been found to be linearly related to the slope of the symmetry energy for nuclear matter, $L_0$~\cite{Warda2012}. Recently, the skin thickness for $^{208}$Pb has been experimentally determined to be 0.15$\pm0.03$ fm through coherent pion photoproduction~\cite{Tarbert2014}. The same value but with an error of $\pm0.02$ fm has been obtained from antiprotonic x-rays while hadron scattering experiments give an average value of 0.17$\pm$0.02 fm \cite{Trz2005}. Figure~\ref{skin} shows predictions for the skin thickness of those nuclei included in the fit with available experimental values.  $\Delta r_{np}$ is plotted against the relative neutron excess, $I=(N-Z)/A$, as defined in Sec.~\ref{nmp}. Higher values of $I$ correspond to neutron-rich nuclei, while symmetric nuclei have $I=0$. In the figure, the gray band is taken from a linear fit of skin thickness experimental data as 
a function of $I$: $\Delta r_{np}(I)=(-0.03\pm0.02)+(0.90\pm0.15)\cdot I$~\cite{Trz2005}.

On the proton-rich side ($I<0$), the mean field models predict a negative value for the skin thickness for $^{36}$Ca and $^{38}$Ca. Towards the neutron-rich side, most values from QMC$\pi$-II are within the error bounds of experiment and while the prediction for $^{208}$Pb is slightly higher than the data taken from antiprotonic x-rays, the QMC result is within the error of the data deduced from hadron scattering experiments. 
\subsubsection{Single-particle states}
\begin{center}
	\begin{figure}[h]
		\centering
		\includegraphics[angle=0,width=1.0\textwidth]{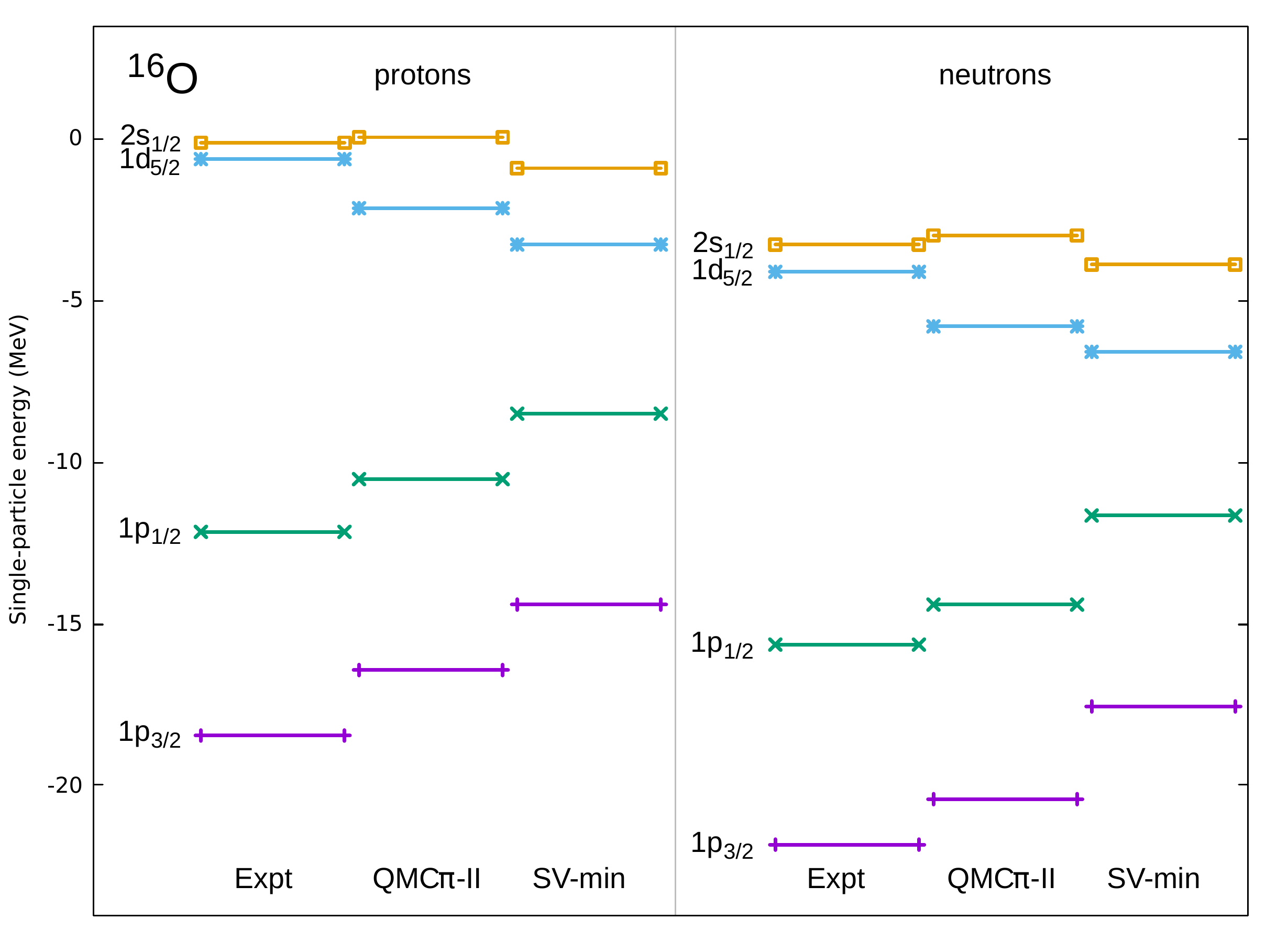}\\
		\caption{Proton and neutron single-particle states for $^{16}$O. Experimental data is taken from Ref.~\cite{Grawe2007} and SV-min values are from Ref.~\cite{frib}. Single-particle levels are shown in different colors and labels are placed before the experimental data for each level.}
		\label{symO}
	\end{figure}
\end{center}

\begin{center}
	\begin{figure}[h]
		\centering
		\includegraphics[angle=0,width=1.0\textwidth]{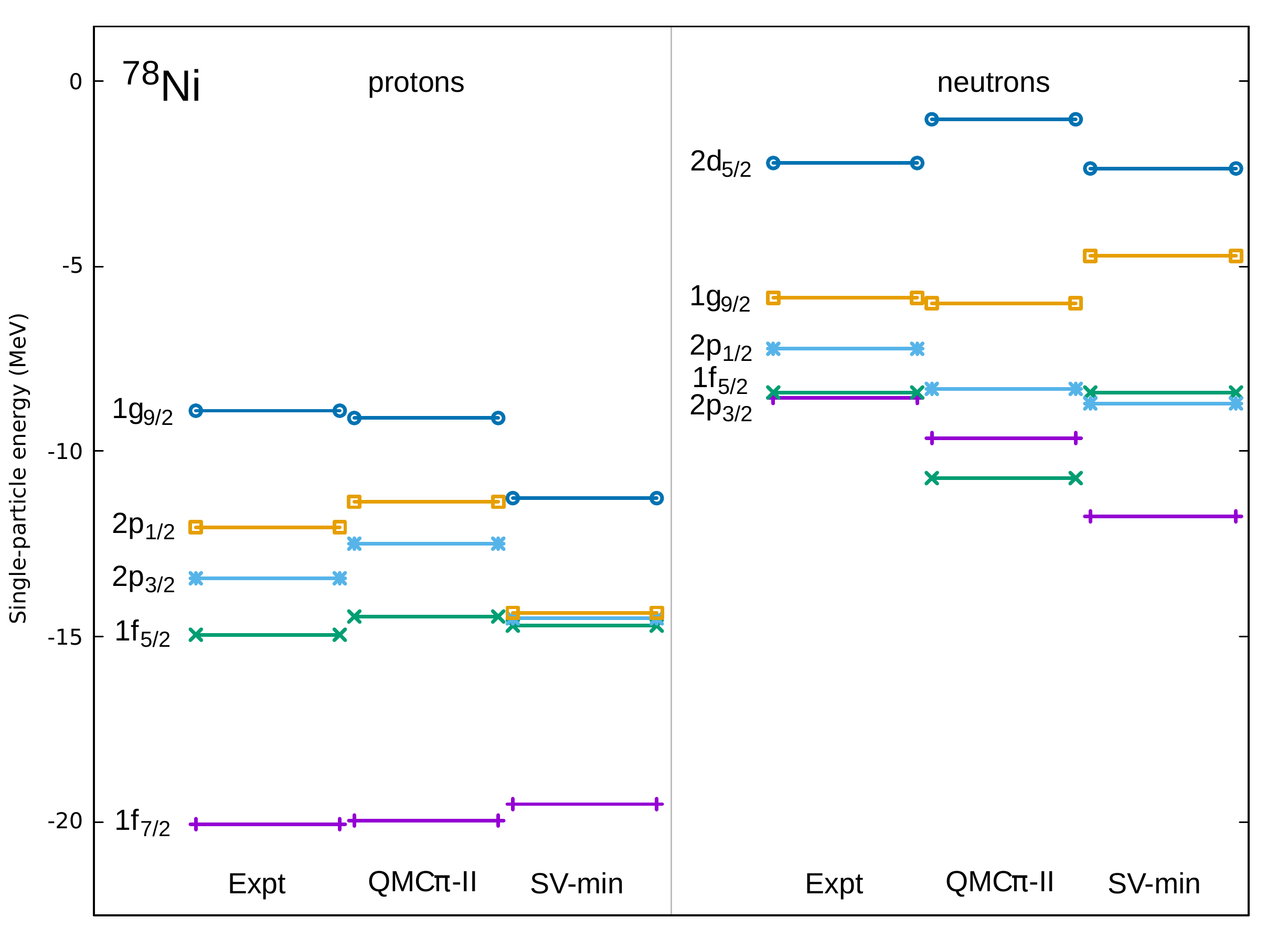}\\
		\caption{Same as in Fig. \ref{symO} but for $^{78}$Ni.}
		\label{symNi}
	\end{figure}
\end{center}

\begin{center}
	\begin{figure}[h]
		\centering
		\includegraphics[angle=0,width=1.0\textwidth]{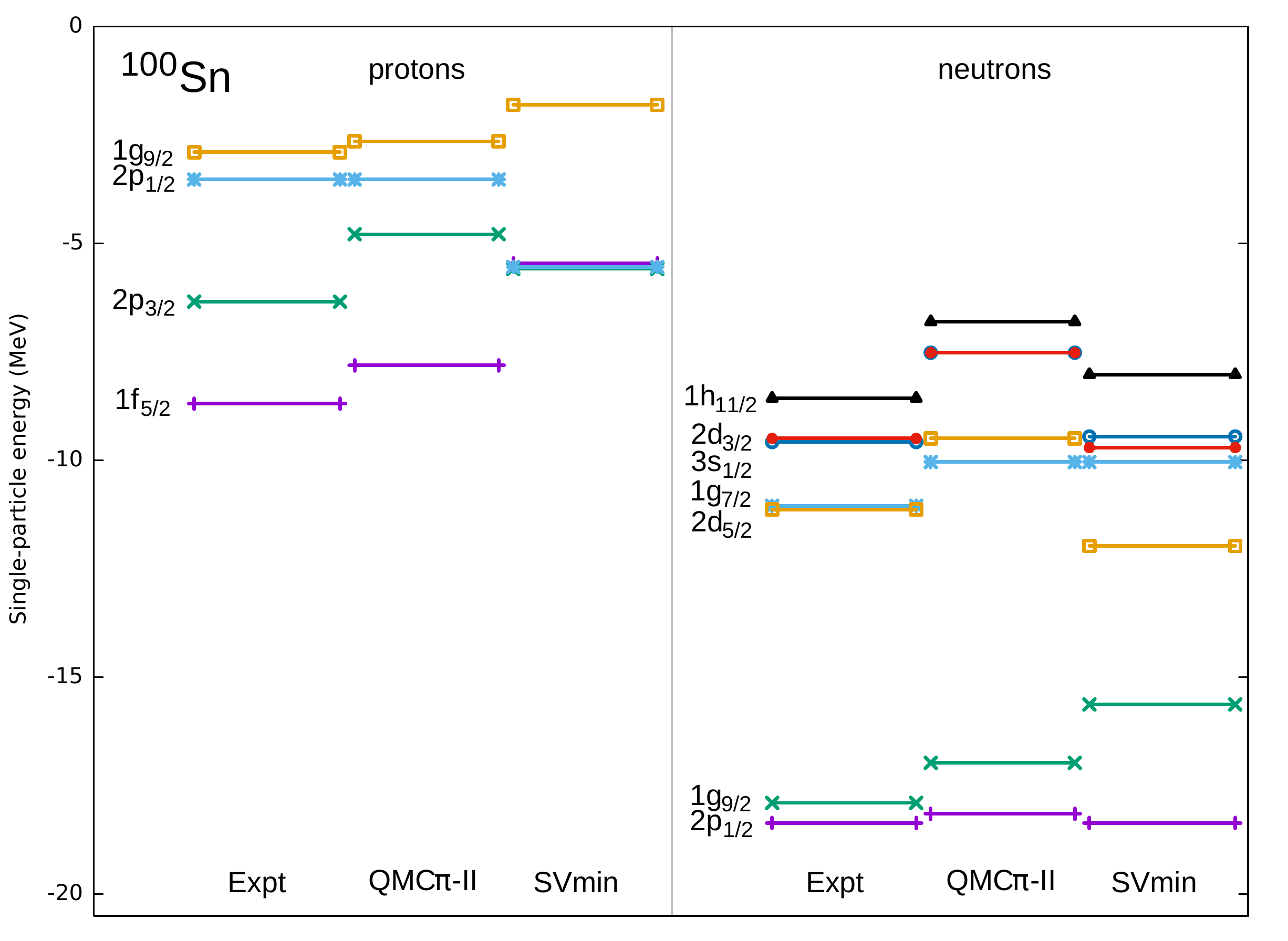}\\
		\caption{Same as in Fig. \ref{symO} but for $^{100}$Sn.}
		\label{symSn100}
	\end{figure}
\end{center}

\begin{center}
	\begin{figure}[h]
		\centering
		\includegraphics[angle=0,width=1.0\textwidth]{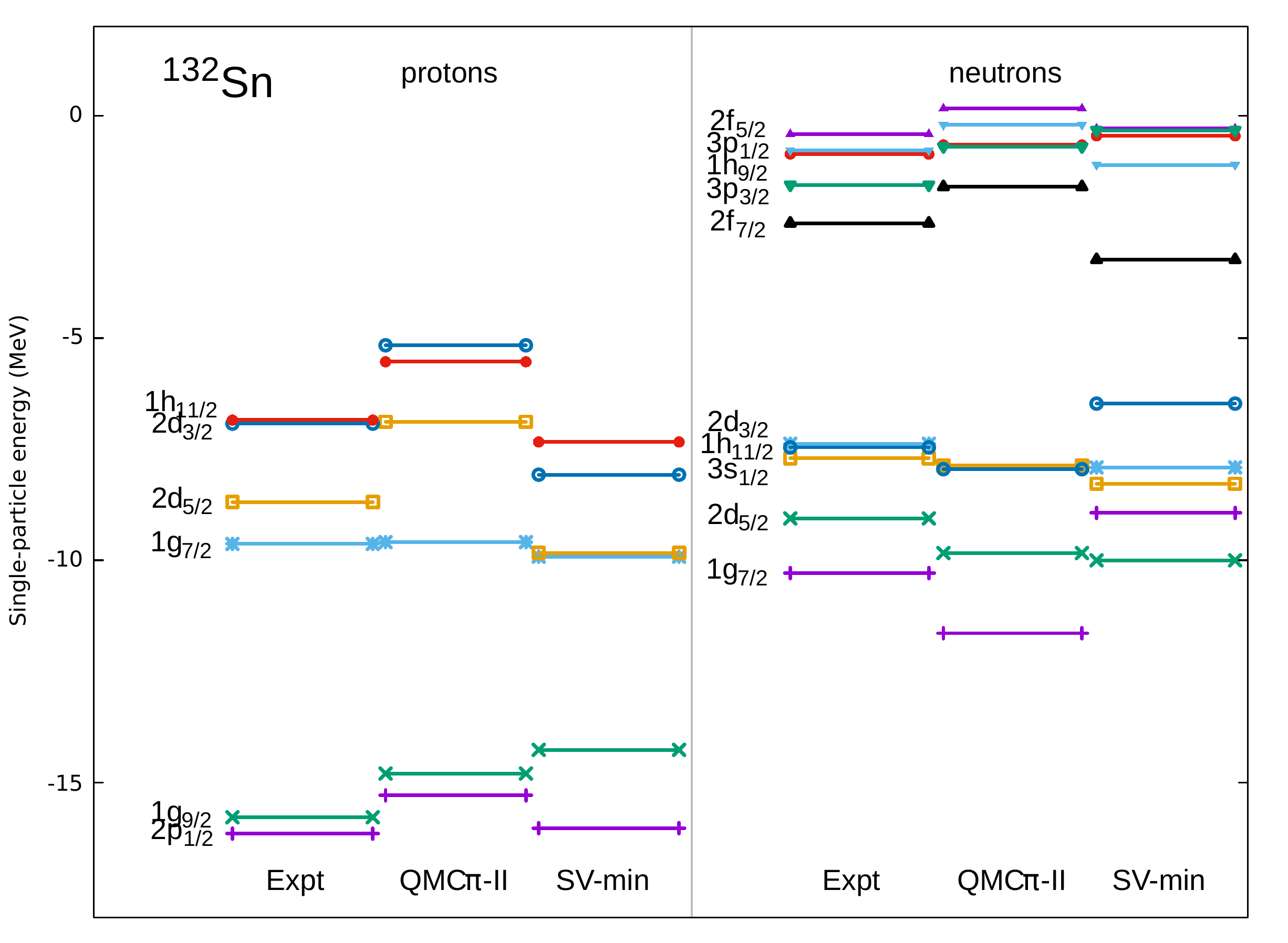}\\
		\caption{Same as in Fig. \ref{symO} but for $^{132}$Sn.}
		\label{symSn132}
	\end{figure}
\end{center}

Figures~\ref{symO} to \ref{symSn132} show neutron and proton single states for doubly magic symmetric $^{16}$O and $^{100}$Sn nuclei as well as for neutron-rich $^{78}$Ni and $^{132}$Sn nuclei. Results for QMC$\pi$-II and SV-min are compared with experimental data from \cite{Grawe2007}. Note that the spin-orbit splittings for some doubly magic nuclei were included in the fitting for SV-min but were left out for the present QMC parameter search.

The calculated values for QMC$\pi$-II for the proton single-particle energies of $^{16}$O, $^{78}$Ni and $^{100}$Sn agree well with experiment, while for $^{132}$Sn some states above the proton Fermi level are more spread out compared to experiment. For the neutron single-particle states, QMC$\pi$-II suffers an inversion of some states in $^{78}$Ni, $^{100}$Sn and $^{132}$Sn compared to the traditional Nilsson scheme. This is also true for some states calculated with SV-min. For instance, for $^{78}$Ni the neutron states $2p_{3/2}$ and $1f_{5/2}$ are inverted for QMC$\pi$-II while states $1f_{5/2}$ and $2p_{1/2}$ are inverted for SV-min. Recall that we saw in Sec.~\ref{obsnot} 
that the QMC results for isotopes near the $N=82$ shell closure had higher residuals for the two-neutron separation energy. This can be seen in the large gap of neutron states in both $^{78}$Ni and $^{132}$Sn, as shown 
in Fig.~\ref{symNi} and Fig.~\ref{symSn132} respectively. 

\subsubsection{Nuclear deformations}
\begin{center}
	\begin{figure}[ht]
		\centering
		\includegraphics[angle=0,width=1.0\textwidth]{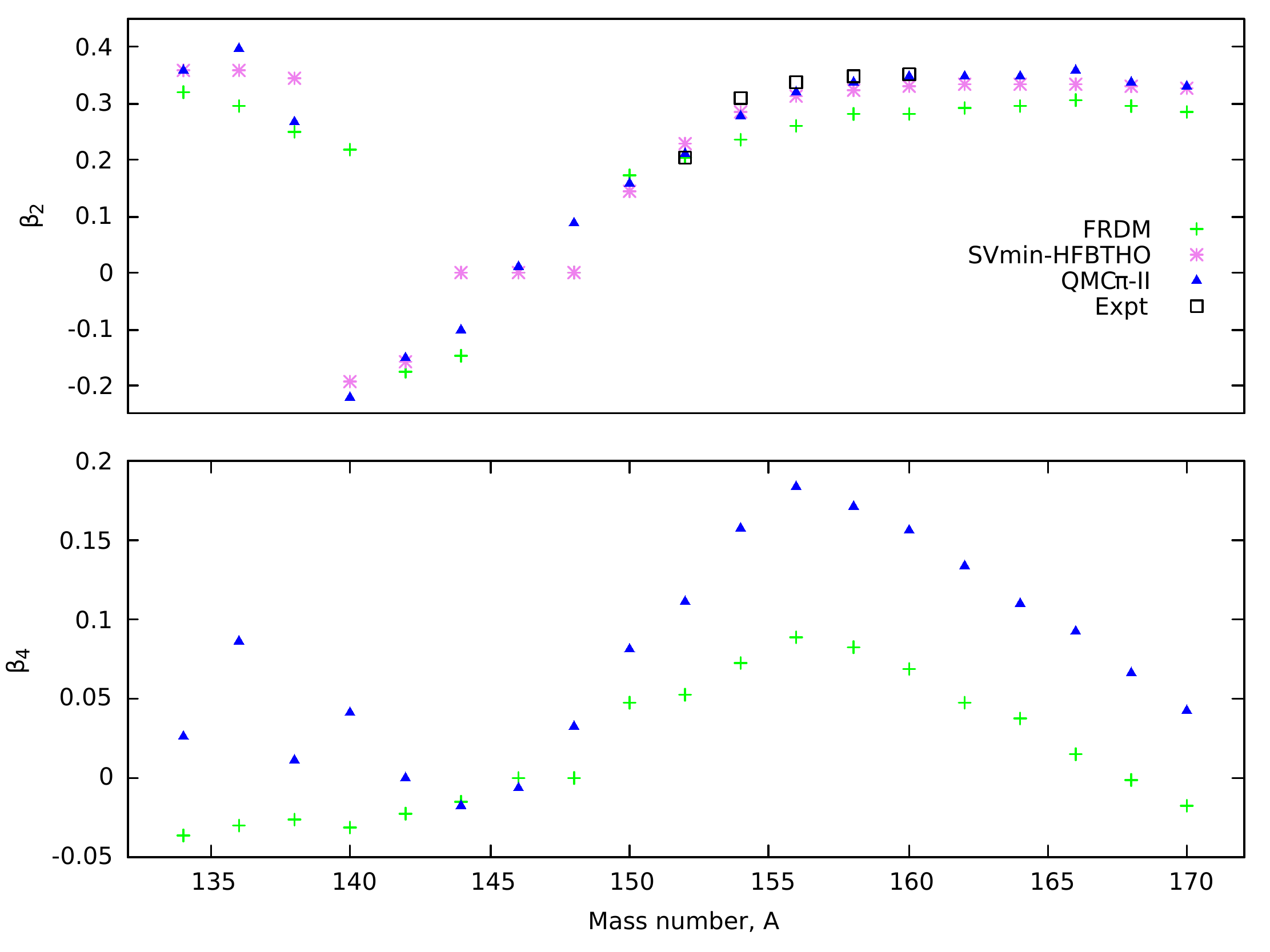}
		\caption{Deformation parameters $\beta_2$ and $\beta_4$ for the Gd isotopes plotted against mass number, $A$. Values for FRDM are taken from Ref.~\cite{Moller2016}, SV-min values are from Ref.~\cite{frib} and the experimental data are from Ref.~\cite{Raman2001}. Plot legend is placed in the top panel.}
		\label{Gddeform}
	\end{figure}
\end{center}

\begin{center}
	\begin{figure}[ht]
		\centering
		\includegraphics[angle=0,width=1.0\textwidth]{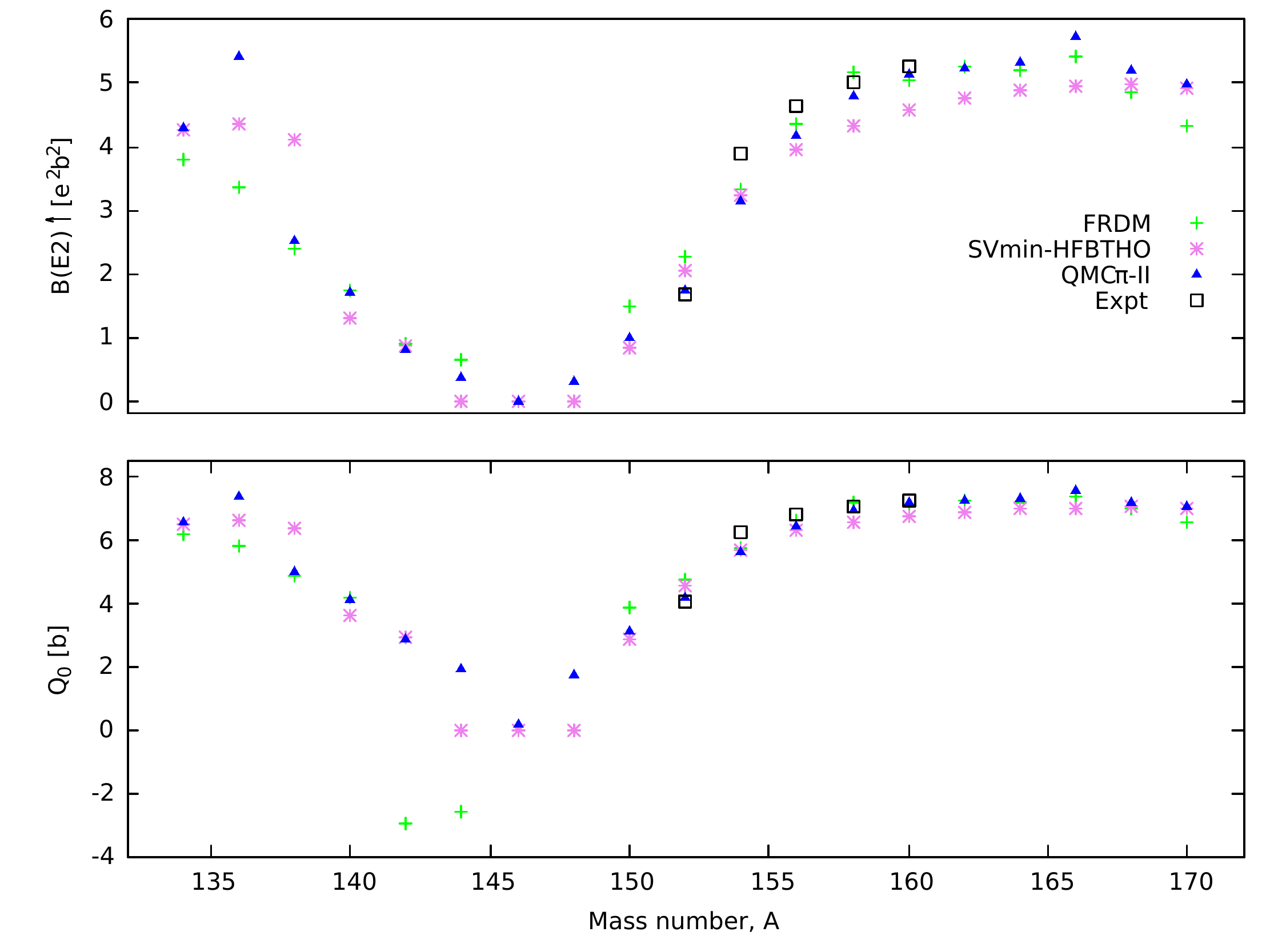}
		\caption{Transition probability, $B(E2)$$\uparrow$, and intrinsic quadrupole moment, $Q_0$, for the Gd isotopes plotted against mass number $A$. $B(E2)$$\uparrow$ and $Q_0$ are computed as in Ref.~\cite{Raman2001}, using 
			the $\beta_2$ values taken from Ref.~\cite{Moller2016} for FRDM and Ref.~\cite{frib} for SV-min. Plot legend is placed in the top panel. }
		\label{Gdmoment}
	\end{figure}
\end{center}

In Refs.~\cite{Kort2010, Kort2012} several deformed nuclei were included in the fitting for UNEDF0 and UNEDF1 parameter sets. Though our current fitting procedure only includes magic nuclei, which are mostly spherical, the final parameter set can be used to extend the calculations to nuclei having deformations. Figure~\ref{Gddeform} shows the performance of QMC in comparison with other models and experiment~\cite{Raman2001} for the deformation parameters, $\beta_2$ and $\beta_4$, of gadolinium ($Z=64$) isotopes. 
The  transition probability $B(E2)$$\uparrow$ from the ground state to the first excited 2$^+$  state and the corresponding intrinsic quadrupole moment, $Q_0$, are added for comparison in Figure~\ref{Gdmoment}. 
As computed in Ref.~\cite{Raman2001}, $B(E2)$$\uparrow$$=\left[(3/4\pi)\beta ZeR_0^2\right]^2$ where $R_0=1.2A^{1/3}$ fm and $Q_0$ is directly related to the transition probability by the expression $Q_0^2=(16\pi/5e^2)B(E2)$$\uparrow$. For FRDM, $Q_0$ is expressed as a function of both $\beta_2$ and $\beta_4$, thus it indirectly provides a check for the $\beta_4$ parameter.\\

The models predict a spherical shape for $^{146}$Gd and prolate shapes for neutron-rich Gd isotopes. We note that computation here is done in constrained case and from Fig. \ref{Gddeform}, isotopes $^{140-144}$Gd are predicted to have oblate shapes as opposed to unconstrained QMC-I results where they were prolate \cite{Stone2016}.  $B(E2)$$\uparrow$ values from QMC$\pi$-II agree well with experimental data where available. As with the calculations of radii in Sec.~\ref{ext}, deformation properties will be the subject of future investigation in the QMC model. 


\subsection{Binding energies of even-even superheavy nuclei}
\label{SHE} 
\begin{center}
	\begin{figure*}[h]
		\centering
		\includegraphics[angle=0,width=1.0\textwidth]{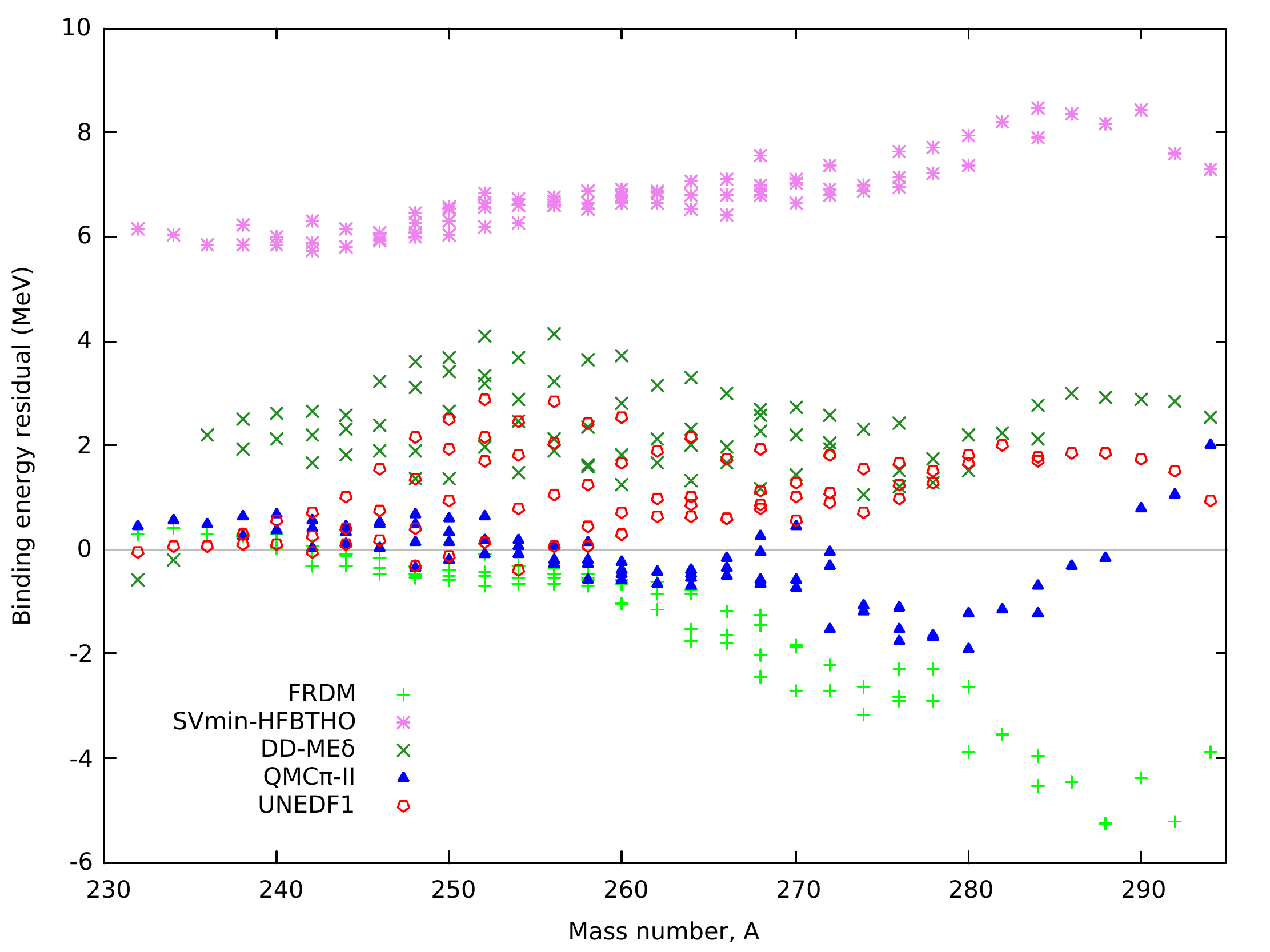}\\
		\caption{SHE binding energy residuals for nuclei with $Z > 96$ plotted against mass number $A$. Also added for comparison are FRDM taken from Ref.~\cite{Moller2016}, SV-min~\cite{Klup2009}, 
			UNEDF1~\cite{Kort2012} and DD-ME$\delta$~\cite{Afa2016}, all taken from Ref.~\cite{frib}. 
			Experimental data used to compute the residuals are from Ref.~\cite{Wang2017}.}
		\label{ebinSHE}
	\end{figure*}
\end{center}

Calculations have also been extended to the superheavy elements with $Z\ge96$, which were not included in the fit. Figure~\ref{ebinSHE} shows excellent QMC predictions for binding energies in comparison with other models. The RMSD for binding energies for these nuclei, calculated with  QMC$\pi$-II, is 0.72 MeV where FRDM and 
DD-ME$\delta$ give 1.9 MeV and 2.5 MeV, respectively. The Skyrme force SV-min and UNEDF1 give 6.8 MeV and 1.4 MeV, respectively. The success of the QMC model in this region was also investigated in the previous version QMC$\pi$-I where $\alpha$ decay energy and deformations were computed~\cite{Stone2017}. Detailed studies of   superheavies in the latest version of the QMC model are currently underway. 

\section{Concluding remarks}
\label{conclude}
Parameter optimisation of the new version of the quark-meson coupling model, QMC$\pi$-II, was carried out using the derivative-free algorithm POUNDeRS. Parameter errors and correlations were presented and the final parameter set was used to calculate various nuclear observables. QMC$\pi$-II produced nuclear matter properties within the acceptable range and showed considerable improvement for the slope of the symmetry energy, as well as for the incompressibility, compared to the values obtained in the previous QMC version. The new QMC parameters were also used to calculate ground state properties of even-even nuclei across the nuclear chart. 
The results were comparable to those of other well known models in the predictions for binding energies, \textit{rms} charge radii and pairing gaps. 
Calculations were extended to other nuclear observables which had not been part of the fit, 
including isotopic and isotonic shifts in energies and radii, skin thickness and single-particle energies, for a number of chosen nuclei. The QMC predictions were shown to be within a similar range to that found with other models. Deformations were also investigated for gadolinium isotopes and are in agreement with available data. The model appears to be particularly effective in the superheavy region, giving an rms binding energy residual of only 0.7 MeV, where other models show higher deviations. Calculations in this region and for unknown SHE up to the drip lines are currently in progress.

In the future, energy calculations will be further investigated for symmetric nuclei as well as in the region of the  $N=50$ and $N=82$ shell closures, where the current QMC residuals are relatively high compared to other regions in the nuclear chart. Charge radii, especially in the region of the mercury and platinum isotopes, and deformation properties of finite nuclei will be analysed further. Odd-mass nuclei will also be studied using the QMC model. In addition, it will be interesting to explore the predictions for nuclei far from stability and currently unknown nuclei up to the proton and neutron 
driplines.\\

\section*{Acknowledgements}

J. R. S. and P. A. M. G. acknowledge with pleasure the support and hospitality of CSSM at the University of Adelaide during visits in the course of this project.

This work was supported by the University of Adelaide and by the Australian Research Council through the 
ARC Centre of Excellence for Particle Physics at the Terascale (CE110001104) and Discovery Projects 
DP150103101 and DP180100497.


\bibliography{Paper1} 
\end{document}